\documentstyle[12pt]{article}
\newlength{\dinwidth}  
\newlength{\dinmargin}
\global\arraycolsep=2pt 

\newcommand{\AmS}{{\protect\the\textfont2
 A\kern-.1667em\lower.5ex\hbox{M}\kern-.125emS}}

\begin{document}
\newcommand{\nn}{\noindent}
\newcommand{\cs}{\mbox{$\clubsuit$}}
\newcommand{\nl}{\nonumber \\}
\newcommand{\hf}{\hfill}
\newcommand{\naive}{na$\ddot{\imath}$ve}
\newcommand {\oa} {\mbox{${\cal O}( \alpha)$}}
\newcommand {\ho} {\mbox{${\cal O}( \alpha^{2})$}}
\hyphenation{brems-strah-lung}
\def\ss{\footnotesize}
\def\SS{\footnotesize}
\def\sss{\scriptscriptstyle}
\def\barp{{\raise.35ex\hbox{${\sss (}$}}---{\raise.35ex\hbox{${\sss )}$}}}
\def\bdbarp{\hbox{$B_d$\kern-1.4em\raise1.4ex\hbox{\barp}}}
\def\bsbarp{\hbox{$B_s$\kern-1.4em\raise1.4ex\hbox{\barp}}}
\def\dbarp{\hbox{$D$\kern-1.1em\raise1.4ex\hbox{\barp}}}
\def\dcp{D^0_{\sss CP}}
\def\dbar{{\overline{D^0}}}
\def\ks{K_{\sss S}}
\newcommand{\xd}{x_d}
\newcommand{\xs}{x_s}
\newcommand{\bd}{B_d^0}
\newcommand{\bdb}{\overline{B_d^0}}
\newcommand{\bs}{B_s^0}
\newcommand{\bsbar}{\overline{B_s^0}}
\newcommand{\bu}{B_u^\pm}
\newcommand{\beq}{\begin{equation}}
\newcommand{\eeq}{\end{equation}}
\newcommand{\absvcb}{\vert V_{cb}\vert}
\newcommand{\absvub}{\vert V_{ub}\vert}
\newcommand{\absvtd}{\vert V_{td}\vert}
\newcommand{\absvts}{\vert V_{ts}\vert}
\newcommand{\abseps}{\vert\epsilon\vert}
\newcommand{\epsp}{\epsilon^\prime/\epsilon}
\newcommand{\fbb}{f^2_{B_d}\hat{B}_{B_d}}
\newcommand{\fbbs}{f^2_{B_s}\hat{B}_{B_s}}
\newcommand{\fbd}{f_{B_d}}
\newcommand{\fbs}{f_{B_s}}
\newcommand{\fds}{f_{D_s}}
\def\rly#1{\mathrel{\raise.3ex\hbox{$#1$\kern-.75em\lower1ex\hbox{$\sim$}}}}
\def\lsim{\rly<}

\def \zpc#1#2#3{{\rm Z.~Phys.} {\bf C#1} (19#2) #3}
\def \plb#1#2#3{{\rm Phys.~Lett.} {\bf B#1} (19#2) #3}
\def \ibj#1#2#3{~#1, (19#2) #3}
\def \prl#1#2#3{{\rm Phys.~Rev.~Lett.} {\bf #1} (19#2) #3}
\def \prd#1#2#3{{\rm Phys.~Rev.} {\bf D#1} (19#2) #3} 
\def \npb#1#2#3{{\rm Nucl.~Phys.} {\bf B#1} (19#2) #3} 
\def\ijmp#1#2#3{{\rm Int.\ J.\ Mod.\ Phys.} {\bf A#1} (19#2) #3}
\def \stone{{\it B Decays}, edited by S. Stone (World Scientific, Singapore,
1994)}

\newread\epsffilein 
\newif\ifepsffileok 
\newif\ifepsfbbfound 
\newif\ifepsfverbose 
\newdimen\epsfxsize 
\newdimen\epsfysize 
\newdimen\epsftsize 
\newdimen\epsfrsize 
\newdimen\epsftmp 
\newdimen\pspoints 
\pspoints=1bp 
\epsfxsize=0pt 
\epsfysize=0pt 
\def\epsfbox#1{\global\def\epsfllx{72}\global\def\epsflly{72}%
 \global\def\epsfurx{540}\global\def\epsfury{720}%
 \def\lbracket{[}\def\testit{#1}\ifx\testit\lbracket
 \let\next=\epsfgetlitbb\else\let\next=\epsfnormal\fi\next{#1}}%
\def\epsfgetlitbb#1#2 #3 #4 #5]#6{\epsfgrab #2 #3 #4 #5 .\\%
 \epsfsetgraph{#6}}%
\def\epsfnormal#1{\epsfgetbb{#1}\epsfsetgraph{#1}}%
\def\epsfgetbb#1{%
%
%
\openin\epsffilein=#1
\ifeof\epsffilein\errmessage{I couldn't open #1, will ignore it}\else
%
%
 {\epsffileoktrue \chardef\other=12
 \def\do##1{\catcode`##1=\other}\dospecials \catcode`\ =10
 \loop
 \read\epsffilein to \epsffileline
 \ifeof\epsffilein\epsffileokfalse\else
%
%
 \expandafter\epsfaux\epsffileline:. \\%
 \fi
 \ifepsffileok\repeat
 \ifepsfbbfound\else
 \ifepsfverbose\message{No bounding box comment in #1; using defaults}\fi\fi
 }\closein\epsffilein\fi}%
%
%
\def\epsfclipstring{}
\def\epsfclipon{\def\epsfclipstring{ clip}}%
\def\epsfclipoff{\def\epsfclipstring{}}%
\def\epsfsetgraph#1{%
 \epsfrsize=\epsfury\pspoints
 \advance\epsfrsize by-\epsflly\pspoints
 \epsftsize=\epsfurx\pspoints
 \advance\epsftsize by-\epsfllx\pspoints
%
%
 \epsfxsize\epsfsize\epsftsize\epsfrsize
 \ifnum\epsfxsize=0 \ifnum\epsfysize=0
 \epsfxsize=\epsftsize \epsfysize=\epsfrsize
 \epsfrsize=0pt
%
%
 \else\epsftmp=\epsftsize \divide\epsftmp\epsfrsize
 \epsfxsize=\epsfysize \multiply\epsfxsize\epsftmp
 \multiply\epsftmp\epsfrsize \advance\epsftsize-\epsftmp
 \epsftmp=\epsfysize
 \loop \advance\epsftsize\epsftsize \divide\epsftmp 2
 \ifnum\epsftmp>0
 \ifnum\epsftsize<\epsfrsize\else
 \advance\epsftsize-\epsfrsize \advance\epsfxsize\epsftmp \fi
 \repeat
 \epsfrsize=0pt
 \fi
 \else \ifnum\epsfysize=0
 \epsftmp=\epsfrsize \divide\epsftmp\epsftsize
 \epsfysize=\epsfxsize \multiply\epsfysize\epsftmp
 \multiply\epsftmp\epsftsize \advance\epsfrsize-\epsftmp
 \epsftmp=\epsfxsize
 \loop \advance\epsfrsize\epsfrsize \divide\epsftmp 2
 \ifnum\epsftmp>0
 \ifnum\epsfrsize<\epsftsize\else
 \advance\epsfrsize-\epsftsize \advance\epsfysize\epsftmp \fi
 \repeat
 \epsfrsize=0pt
 \else
 \epsfrsize=\epsfysize
 \fi
 \fi
%
%
 \ifepsfverbose\message{#1: width=\the\epsfxsize, height=\the\epsfysize}\fi
 \epsftmp=10\epsfxsize \divide\epsftmp\pspoints
 \vbox to\epsfysize{\vfil\hbox to\epsfxsize{%
 \ifnum\epsfrsize=0\relax
 \includegraphics{#1}%
 \else
 \epsfrsize=10\epsfysize \divide\epsfrsize\pspoints
 \includegraphics{#1}%
 \fi
 \hfil}}%
\global\epsfxsize=0pt\global\epsfysize=0pt}%
%
%
 {\catcode`\%=12 \global\let\epsfpercent=
%
%
\long\def\epsfaux#1#2:#3\\{\ifx#1\epsfpercent
 \def\testit{#2}\ifx\testit\epsfbblit
 \epsfgrab #3 . . . \\%
 \epsffileokfalse
 \global\epsfbbfoundtrue
 \fi\else\ifx#1\par\else\epsffileokfalse\fi\fi}%
%
%
\def\epsfempty{}%
\def\epsfgrab #1 #2 #3 #4 #5\\{%
\global\def\epsfllx{#1}\ifx\epsfllx\epsfempty
 \epsfgrab #2 #3 #4 #5 .\\\else
 \global\def\epsflly{#2}%
 \global\def\epsfurx{#3}\global\def\epsfury{#4}\fi}%
%
%
\def\epsfsize#1#2{\epsfxsize}
%
%
\let\epsffile=\epsfbox
\def\att{t \bar{t}}
\def\app{p \bar{p}}
\def\rts{\sqrt{s}}
\def\mt{m_t}
\def\mb{m_b}
\def\mc{m_c}
\newcommand{\bksgam}{\ $B \to K^*+ \gamma$}
\newcommand{\brogam}{\ $B \to \rho+ \gamma$}
\def\BDSl{B \to D^* \ell \nu_\ell}
\def\vdvp{v \cdot v^\prime}
\def\xiaoo{\xi_{A_1}(\vdvp =1 )}
\def\Vbc{V_{cb}}
\newcommand{\Tosc}{T_{osc}}
\newcommand{\sqrts}{\sqrt{s}}
\newcommand{\bg}{\beta \gamma}
\newcommand{\xds}{x_i}
\newcommand{\Ds}{D_s^\pm}
\newcommand{\bb}{B^0 B^0}
\newcommand{\barbar}{{\overline{B^0}}\thinspace{\overline{B^0}}}
\newcommand{\barb}{B^0 {\overline{B^0}}}
\newcommand{\bbar}{$B^0$--${\overline{B^0}}$}
\newcommand{\Deltat}{\Delta t}
\newcommand{\delt}{\delta t}
\newcommand{\delmd}{\Delta M_d}
\newcommand{\delms}{\Delta M_s}
\newcommand{\ps}{10^{-12} s}
\newcommand{\zbbar}{Z^0 \to b {\overline{b}}}
\newcommand{\eebbx}{$e^+ e^- \to B {\overline{B}} X$}
\newcommand{\pbpbbx}{$p{\overline{p}} \to B {\overline{B}} X$}
\newcommand{\kkbar}{$K^0$--${\overline{K^0}}$}
\newcommand{\bdbdbar}{$B_d^0$--${\overline{B_d^0}}$}
\newcommand{\bsbsbar}{$B_s^0$--${\overline{B_s^0}}$}
\newcommand{\as}{\mbox{$\alpha_{\displaystyle s}$}}
\newcommand{\aso}{\mbox{$O(\alpha_{\displaystyle s})$}}
\newcommand{\ass}{\mbox{$O(\alpha_{\displaystyle s}^2)$}}
\newcommand{\asq}{\mbox{$\alpha_{\displaystyle s}(Q^2)$}}
\newcommand{\ee}{\mbox{$e^+e^-$}}
\newcommand{\cc}{\mbox{$c {\overline{c}}$}}
\newcommand{\qq}{\mbox{$q {\overline{q}}$}}
\newcommand{\jp}{\mbox{$J/\Psi$}}
\newcommand{\lqc}{\Lambda_{QCD}}
\newcommand{\pmi}{{\not{p}}_{\perp}}
\newcommand{\set}{\sum E_{\perp}}
\newcommand{\ptr}{p_{\perp}}
\newcommand{\sww}{\sin^2{\theta_W}}
\newcommand{\sw}{\sin{\theta_W}}
\begin{flushright}
DESY 99-042 \\
UdeM-GPP-TH-99-58\\
March 1999\\
\end{flushright}
\begin{center}
{\Large \bf
\centerline{Profiles of the Unitarity Triangle and CP-Violating}
 \centerline{Phases in the Standard Model and Supersymmetric Theories}}
\vspace*{1.5cm}
{\large A.~Ali} \vskip0.2cm
Deutsches Elektronen Synchrotron DESY, Hamburg \\
\vspace*{0.3cm} \centerline{ and} \vspace*{0.3cm} {\large D.~London}
\vskip0.2cm Laboratoire Ren\'e J.-A. L\'evesque, Universit\'e de
Montr\'eal, C.P. 6128, succ.\ 
centre-ville, Montr\'eal, QC, Canada H3C 3J7 \\
\vskip0.5cm 
{\Large Abstract\\} 
\vskip3truemm
\parbox[t]{\textwidth}{ 
\indent 
We report on a comparative study of the profile of the CKM unitarity
triangle, and the resulting CP asymmetries in $B$ decays, in the
standard model and in several variants of the minimal supersymmetric
standard model (MSSM), characterized by a single phase in the quark
flavour mixing matrix. The supersymmetric contributions to the mass
differences $\Delta M_d$, $\Delta M_s$ and to the CP-violating
quantity $\abseps$ are, to an excellent approximation, equal to each
other in these theories, allowing for a particularly simple way of
implementing the resulting constraints on the elements of $V_{CKM}$
from the present knowledge of these quantities. Incorporating the
next-to-leading-order corrections and applying the current direct and
indirect constraints on the supersymmetric parameters, we find that
the predicted ranges of $\sin 2 \beta$ in the standard model and in
MSSM models are very similar. However, precise measurements at
$B$-factories and hadron machines may be able to distinguish these
theories in terms of the other two CP-violating phases $\alpha$ and
$\gamma$. This is illustrated for some representative values of the
supersymmetric contributions in $\Delta M_d$, $\Delta M_s$ and
$\abseps$.
}
\end{center}
\thispagestyle{empty}
\newpage  
\setcounter{page}{1}
\textheight 23.0 true cm

\section{Introduction}

Within the standard model (SM), CP violation is due to the presence of
a nonzero complex phase in the Cabibbo-Kobayashi-Maskawa (CKM) quark
mixing matrix $V$ \cite{CKM}. A particularly useful parametrization of
the CKM matrix, due to Wolfenstein \cite{Wolfenstein}, follows from
the observation that the elements of this matrix exhibit a hierarchy
in terms of $\lambda$, the Cabibbo angle. In this parametrization the
CKM matrix can be written approximately as
\beq
V \simeq \left(\matrix{
 1-{1\over 2}\lambda^2 & \lambda
 & A\lambda^3 \left( \rho - i\eta \right) \cr
 -\lambda ( 1 + i A^2 \lambda^4 \eta )
& 1-{1\over 2}\lambda^2 & A\lambda^2 \cr
 A\lambda^3\left(1 - \rho - i \eta\right) & -A\lambda^2 & 1 \cr}\right)~.
\label{CKM}
\eeq
The allowed region in $\rho$--$\eta$ space can be elegantly displayed
using the so-called unitarity triangle (UT). The unitarity of the CKM
matrix leads to the following relation:
\beq
V_{ud} V_{ub}^* + V_{cd} V_{cb}^* + V_{td} V_{tb}^* = 0~.
\eeq
Using the form of the CKM matrix in Eq.~(\ref{CKM}), this can be recast as
\beq
\label{trianglerel}
\frac{V_{ub}^*}{\lambda V_{cb}} + \frac{V_{td}}{\lambda V_{cb}} = 1~,
\eeq
which is a triangle relation in the complex plane (i.e.\ $\rho$--$\eta$
space), illustrated in Fig.~\ref{triangle}. Thus, allowed values of $\rho$
and $\eta$ translate into allowed shapes of the unitarity triangle.

\begin{figure}
\vskip -1.0truein
\centerline{\epsfxsize 3.5 truein \epsfbox {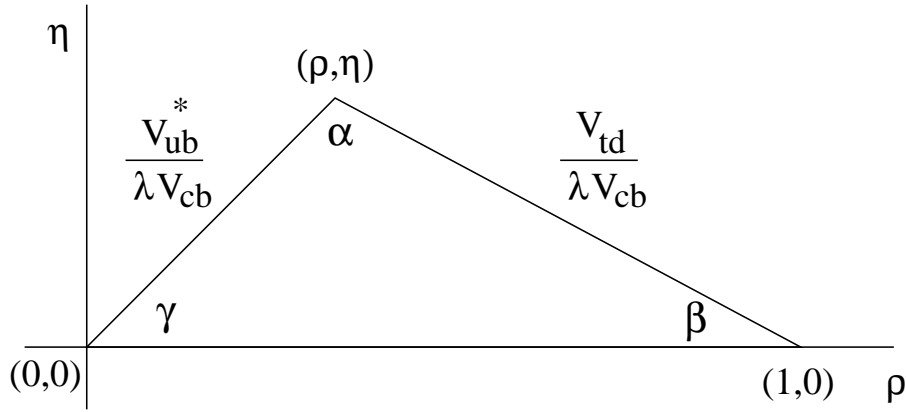}}
\vskip -1.2truein
\caption{The unitarity triangle. The angles $\alpha$, $\beta$ and $\gamma$
can be measured via CP violation in the $B$ system.}
\label{triangle}
\end{figure}

Constraints on $\rho$ and $\eta$ come from a variety of sources. Of
the quantities shown in Fig.~\ref{triangle}, $|V_{cb}|$ and $|V_{ub}|$
can be extracted from semileptonic $B$ decays, while $|V_{td}|$ is
probed in \bdbdbar\ mixing. The interior CP-violating angles $\alpha$,
$\beta$ and $\gamma$ can be measured through CP asymmetries in $B$
decays \cite{BCPasym}. Additional constraints come from CP violation
in the kaon system ($\abseps$), as well as \bsbsbar\ mixing.

In light of the fact that the $B$-factory era is almost upon us, one
of the purposes of this paper is to update the profile of the
unitarity triangle within the SM using the latest experimental data.
This analysis is done at next-to-leading-order (NLO) precision, taking
into account the state-of-the-art calculations of the hadronic matrix
elements from lattice QCD and available data. This therefore provides
a theoretically-robust overview of the SM expectations for the allowed
values of the CP-violating phases, as well as their correlations.

These CP phases are expected to be measured in the very new future.
If Nature is kind, the unitarity triangle, as constructed from direct
measurements of $\alpha$, $\beta$ and $\gamma$, will be inconsistent
with that obtained from independent measurements of the sides. If this
were to happen, it would be clear evidence for the presence of physics
beyond the SM, and would be most exciting.

One type of new physics which has been extensively studied is
supersymmetry (SUSY). There are a number of hints suggesting that SUSY
might indeed be around the corner. One example is gauge-coupling
unification: a supersymmetric grand unified theory does better than
its non-supersymmetric counterpart. (How compelling these hints are
depends on one's point of view.) Partly motivated by this success, but
mostly on theoretical grounds, a great deal of effort has gone into a
systematic study of the pattern of flavour violation in SUSY, where a
number of flavour-changing neutral-current processes in $K$ and $B$
decays have been studied. In this paper we investigate the profile of
the unitarity triangle in supersymmetric models. In particular, we
explore the extent to which SUSY can be discovered through
measurements of the sides and angles of the unitarity triangle.

If new physics (of any type) is present, the principal way in which it
can enter is via new contributions, possibly with new phases, to
\kkbar, \bdbdbar\ and \bsbsbar\ mixing \cite{NPBmixing}. The decay
amplitudes, being dominated by virtual $W$ exchange, remain
essentially unaffected by new physics. Thus, even in the presence of
new physics, the measured values of $|V_{cb}|$ and $|V_{ub}|$
correspond to their true SM values, so that two sides of the UT are
unaffected (see Fig.~\ref{triangle}). However, the third side, which
depends on $|V_{td}|$, will in general be affected by new physics.
Furthermore, the measurements of $\abseps$ and \bsbsbar\ mixing,
which provide additional constraints on the UT, will also be affected.
Therefore, if new physics is present, the allowed region of the UT, as
obtained from current experimental data, may not correspond to the
true (SM) allowed region.

The CP angles $\alpha$ and $\beta$ are expected to be measured via
CP-violating asymmetries in $\bd(t) \to \pi^+\pi^-$ and $\bd(t) \to
J/\Psi \ks$, respectively \cite{BCPasym}. If there is new physics in
\bdbdbar\ mixing, {\it with new phases}, then these measurements will
be affected. On the other hand, if there are no new phases, then the
measurements will probe the true SM values. The third angle $\gamma$
can be measured in a variety of ways. If it is obtained from a CP
asymmetry involving a neutral $B$ meson ($\bd$ or $\bs$), then it may
be affected by new physics. However, if it is measured via charged $B$
decays, then the measured value will correspond to the true (SM)
value.

The most general SUSY models allow for the presence of new phases in
the couplings of supersymmetric and ordinary particles. In such
models the new phases are essentially unconstrained, so that the
measured CP phases can be greatly shifted from their SM values. If
this is the case, then the new physics will be relatively easy to
find: the unitarity triangle constructed from the measurements of
$\alpha$, $\beta$ and $\gamma$ will be considerably different from
that constructed from measurements of the sides. However, precisely
because these new phases are unconstrained, these models lack
predictivity. All that one can say is that there are regions of
parameter space in which large effects are possible. (Indeed, this is
true for any model of new physics which allows for new phases.
Examples include models with four generations, $Z$-mediated
flavour-changing neutral currents, flavour-changing neutral scalars,
etc.\ \cite{Bnewphysics}.)

In part because of this, most of the theoretical attention has
focussed on SUSY models which, though less general, have considerably
more predictive power. The most theoretically-developed model is the
minimal supersymmetric standard model (MSSM). Although this is often
referred to as a single model, in point of fact there are several
variants of the MSSM. Among these is the scenario of minimal
supersymmetric flavour violation \cite{CDGG98}, which involves, in
addition to the SM degrees of freedom, charged Higgs bosons, a light
stop (assumed right-handed) and a light chargino, with all other
degrees of freedom assumed heavy and hence effectively integrated out.
This scenario can be embedded in supergravity (SUGRA) models with
gauge-mediated supersymmetry breaking, in which the first two squark
generations and the gluinos are assumed heavy. Regardless of which
variant is used, the key point for our purposes is that there are no
new phases in the couplings -- although there are many new
contributions to meson mixing, all are proportional to the same
combination of CKM matrix elements as found in the SM. As explained
above, in this class of models measurements of the CP phases will
yield the true SM values for these quantities. However, measurements
of meson mixing will be affected by the presence of this new physics.

In this minimal SUSY scenario, NLO calculations for $\delmd$, $\delms$
and $\abseps$ can be extracted from the work of Krauss and Soff
\cite{KS98}. Also, NLO corrections to the decay $B \to X_s \gamma$
have been worked out by Ciuchini et al.\ \cite{CDGG98}. We make use
of this work and present the profile of the unitarity triangle and
CP-violating phases in this scenario, at the NLO precision. A
particularly nice feature is that the SUSY contributions to \kkbar,
\bdbdbar\ and \bsbsbar\ mixing have the same form. Thus, as far as the
unitarity triangle is concerned, the various SUSY models can be
distinguished by a single parameter, $f$. This simplifies the analysis
considerably.

We note that bits and pieces of such an analysis are already present
in the literature. However, a theoretically-consistent analysis of the
CKM unitarity triangle and the CP-violating phases, taking into
account all constraints, has not been performed yet, to the best of
our knowledge.

With this analysis, one can explore the extent to which the presence
of minimal supersymmetry can be discovered through the precision
measurements of the unitarity triangle which will be undertaken by
experiments at the $B$-factories and hadron colliders. As we will see,
the profiles of the unitarity triangle in the SM and in MSSM models
are similar. However, precise measurements may be able to distinguish
them.

In Section 2, we discuss the profile of the unitarity triangle within
the SM. We describe the input data used in the fits and present the
allowed region in $\rho$--$\eta$ space, as well as the
presently-allowed ranges for the CP angles $\alpha$, $\beta$ and
$\gamma$. We turn to supersymmetric models in Section 3. We review
several variants of the MSSM, in which the new phases are essentially
zero.  We also discuss the NLO corrections in such models and show
that the SUSY contributions to \kkbar, \bdbdbar\ and \bsbsbar\ mixing
are of the same form and can be characterized by a single parameter
$f$. We compare the profile of the unitarity triangle in SUSY models,
for various values of $f$, with that of the SM. We conclude in Section
4.

\section{Unitarity Triangle: SM Profile}

\subsection{Input Data}

The CKM matrix as parametrized in Eq.~(\ref{CKM}) depends on four
parameters: $\lambda$, $A$, $\rho$ and $\eta$. We summarize below the
experimental and theoretical data which constrain these CKM
parameters.

\begin{itemize}
  
\item {$\vert V_{us} \vert $}: We recall that $\vert
  V_{us}\vert$ has been extracted with good accuracy from $K\to\pi
  e\nu$ and hyperon decays \cite{PDG98} to be
\beq
\vert V_{us}\vert=\lambda=0.2196\pm 0.0023~.
\eeq
In our fits, we ignore the small error on $\lambda$.

\item {$ \absvcb $}: The determination of $\absvcb$ from
  inclusive and exclusive $B$ decays has been studied in a number of
  papers \cite{neubert97,czarnecki97,shifman95,Parodiconf98,Mele98}.
  The number used here is taken from the Particle Data Group
  compilation \cite{PDG98}:
\beq 
 \vert V_{cb} \vert = 0.0395 \pm 0.0017~,
\label{vcbnow}
\eeq
yielding
\beq
A = 0.819 \pm 0.035~.
\label{Avalue}
\eeq

\item {$  |V_{ub}/V_{cb}| $}: The knowledge of the CKM matrix
  element ratio $|V_{ub}/V_{cb}|$ is based on the analysis of the
  end-point lepton energy spectrum in semileptonic decays $B \to X_{u}
  \ell \nu_\ell$ and the measurement of the exclusive semileptonic
  decays $B \to (\pi, \rho) \ell \nu_\ell$. Present measurements in
  both the inclusive and exclusive modes are compatible with
  \cite{Parodiconf98}:
\beq
\left\vert \frac{V_{ub}}{V_{cb}} \right\vert = 0.093\pm 0.014~.
\label{vubvcbn}
\eeq
This gives
\beq
\sqrt{\rho^2 + \eta^2} = 0.423 \pm 0.064~.
\eeq

\item {$ \abseps, \hat{B}_K$}: The experimental value of
$\abseps$ is \cite{PDG98}:
\beq
\abseps = (2.280\pm 0.013)\times 10^{-3}~.
\eeq
In the standard model, $\abseps$ is essentially proportional to the
imaginary part of the box diagram for \kkbar\ mixing and is given by
\cite{Burasetal}
\begin{eqnarray}
\abseps &=& \frac{G_F^2f_K^2M_KM_W^2}{6\sqrt{2}\pi^2\Delta M_K}
\hat{B}_K\left(A^2\lambda^6\eta\right)
\bigl(y_c\left\{\hat{\eta}_{ct}f_3(y_c,y_t)-\hat{\eta}_{cc}\right\}
 \nonumber \\
&~& ~~~~~~~~~~~~~~+ 
~\hat{\eta}_{tt}y_tf_2(y_t)A^2\lambda^4(1-\rho)\bigr), 
\label{eps}
\end{eqnarray}
where $y_i\equiv m_i^2/M_W^2$, and the functions $f_2$ and $f_3$ are:
\begin{eqnarray}
f_2(x)&=&\frac{1}{4} +\frac{9}{4}\frac{1}{(1-x)} 
-\frac{3}{2}\frac{1}{(1-x)^2} -\frac{3}{2} \frac{x^2\ln x}{(1-x)^3} ~,
\nonumber\\
f_3(x,y)&=& \ln\frac{x}{y} - \frac{3y}{4(1-y)}\left(1 + \frac{y}{1-y} \ln 
y \right) ~.
\label{f2f3}
\end{eqnarray}
(The above form for $f_3(x,y)$ is an approximation, obtained in the
limit $x \ll y$. For the exact expression, see Ref.~\cite{InamiLim}.)
Here, the $\hat{\eta}_i$ are QCD correction factors, calculated at
next-to-leading order in Refs.~\cite{HN94} ($\hat{\eta}_{cc}$),
\cite{etaB} ($\hat{\eta}_{tt}$) and \cite{HN95} ($\hat{\eta}_{ct}$).
The theoretical uncertainty in the expression for $\abseps$ is in the
renormalization-scale independent parameter $\hat{B}_K$, which
represents our ignorance of the hadronic matrix element $\langle K^0
\vert {({\overline{d}}\gamma^\mu (1-\gamma_5)s)}^2 \vert
{\overline{K^0}}\rangle$. Recent calculations of $\hat{B}_K$ using
lattice QCD methods are summarized in Refs.~\cite{Draper98,Sharpe98},
yielding
\begin{equation}
 \hat{B}_K=0.94 \pm 0.15 .
\label{BKrange}
\end{equation}

\item {$ \Delta M_d, \fbb$}: The present world average for
  $\Delta M_d$ is \cite{Alexander98}
\beq
\Delta M_d = 0.471 \pm 0.016~(ps)^{-1} ~.
\eeq
The mass difference $\Delta M_d$ is calculated from the \bdbdbar\ box
diagram. Unlike the kaon system, where the contributions of both the
$c$- and the $t$-quarks in the loop are important, this diagram is
dominated by $t$-quark exchange:
\beq
\label{bdmixing}
\Delta M_d = \frac{G_F^2}{6\pi^2}M_W^2M_B\left(\fbb\right)\hat{\eta}_B y_t
f_2(y_t) \vert V_{td}^*V_{tb}\vert^2~, \label{xd}
\eeq
where, using Eq.~(\ref{CKM}), $\vert V_{td}^*V_{tb}\vert^2=
A^2\lambda^{6} [\left(1-\rho\right)^2+\eta^2]$. Here, $\hat{\eta}_B$
is the QCD correction. In the fits presented in Ref.~\cite{AL96} we used
the value $\hat{\eta}_B=0.55$, calculated in the $\overline{MS}$
scheme, following Ref.~\cite{etaB}. Consistency requires that the top
quark mass be rescaled from its pole (mass) value of $\mt =175 \pm 5$
GeV to the value $\overline{\mt}(\mt(pole))=165 \pm 5$ GeV in the
$\overline{MS}$ scheme. We shall ignore the slight dependence of 
$\hat{\eta}_B$ on $\overline{\mt}(\mt(pole))$ in the range given here. 

For the $B$ system, the hadronic uncertainty is given by $\fbb$,
analogous to $\hat{B}_K$ in the kaon system, except that in this case
$\fbd$ has not been measured. Present estimates of this quantity are
summarized in Refs.~\cite{Draper98,Sharpe98}, yielding $\fbd
\sqrt{\hat{B}_{B_d}} =(190 \pm 23)$ MeV in the quenched approximation.
The effect of unquenching is not yet understood completely. Taking the
MILC collaboration estimates of unquenching would increase the central
value of $\fbd \sqrt{\hat{B}_{B_d}}$ by $21$ MeV \cite{MILC98}. In our
fits, we have taken
\begin{equation}
\fbd \sqrt{\hat{B}_{B_d}} = 215 \pm 40 ~\mbox{MeV}~,
\label{FBrange}
\end{equation}
which is a fairly conservative estimate of the present theoretical
error on this quantity.

\item {$ \Delta M_s, \fbbs$}: Mixing in the \bsbsbar\ system
  is quite similar to that in the \bdbdbar\ system. The \bsbsbar\ box
  diagram is again dominated by $t$-quark exchange, and the mass
  difference between the mass eigenstates $\delms$ is given by a
  formula analogous to that of Eq.~(\ref{xd}):
\beq
\delms = \frac{G_F^2}{6\pi^2}M_W^2M_{B_s}\left(\fbbs\right)
\hat{\eta}_{B_s} y_t f_2(y_t) \vert V_{ts}^*V_{tb}\vert^2~.
\label{xs}
\eeq
Using the fact that $\vert V_{cb}\vert=\vert V_{ts}\vert$ (Eq.~(\ref{CKM})),
it is clear that one of the sides of the unitarity triangle, $\vert
V_{td}/\lambda V_{cb}\vert$, can be obtained from the ratio of $\delmd$ and
$\delms$,
\beq
\frac{\delms}{\delmd} =
 \frac{\hat{\eta}_{B_s}M_{B_s}\left(\fbbs\right)}
{\hat{\eta}_{B_d}M_{B_d}\left(\fbb\right)}
\left\vert \frac{V_{ts}}{V_{td}} \right\vert^2.
\label{xratio}
\eeq
The only real uncertainty in this quantity is the ratio of hadronic
matrix elements $\fbbs/\fbb$. It is now widely accepted that the ratio
$\xi_s \equiv (f_{B_s} \sqrt{\hat{B}_{B_s}}) /
(f_{B_d}\sqrt{\hat{B}_{B_d}})$ is probably the most reliable of the
lattice-QCD estimates in $B$ physics, and the present estimate is
\cite{Draper98,Sharpe98}:
\beq
\label{xirange}
\xi_s=1.14\pm 0.06 ~.
\eeq

The present lower bound on $\Delta M_s$ is: $\Delta M_s > 12.4
~\mbox{(ps)}^{-1}$ (at $95\%$ C.L.) \cite{Parodiconf98}. This bound
has been established using the so-called ``amplitude method''
\cite{Moser97}. This method is also the best procedure for including
information about \bsbsbar\ mixing in the fit, and works as follows.
Given a meson which at $t=0$ was pure $\bs$, the probability for it to
be detected as a $\bs$ at time $t$ is
\beq
P = {1 \over 2\tau} e^{-t/\tau} \left( 1 + \cos \Delta M_s t \right),
\eeq
while the probability for it to be detected as a $\bsbar$ is
\beq
P = {1 \over 2\tau} e^{-t/\tau} \left( 1 - \cos \Delta M_s t \right).
\eeq
In the amplitude method, one introduces the amplitude ${\cal A}$ and
writes the oscillations terms as $\left( 1 \pm {\cal A} \cos \Delta
  M_s t \right)$. One then measures the value of ${\cal A}$, along
with its error $\sigma_{\cal A}$ assuming various values of $\Delta
M_s$. For a given value of $\Delta M_s$, if ${\cal A}$ is compatible
with 0, one concludes that there is no visible oscillation at this
frequency; if ${\cal A}$ is compatible with 1, one concludes that an
oscillation was observed at this frequency.

The experimental data consists of measured values of ${\cal A}$ and
$\sigma_{\cal A}$ for various values of $\Delta M_s$. To include this
data in the fit, for each set of free parameters $(A,\rho,\eta,\xi_s)$
we calculate the value of $\Delta M_s$ and find the corresponding
experimental values of ${\cal A}$ and $\sigma_{\cal A}$. Since a
nonzero value of $\Delta M_s$ implies that there is \bsbsbar\ mixing,
theoretically one should have ${\cal A} = 1$. For this set of
parameters we therefore add to the global $\chi^2$ a factor
\beq
\left( {{\cal A} - 1 \over \sigma_{\cal A}} \right)^2 ~.
\eeq
(There is also a similar factor which takes into account the deviation
of $\xi_s$ from its central value (Eq.~(\ref{xirange})).)

\end{itemize}

The data used in our fits are summarized in Table \ref{datatable}.
The quantities with the largest errors are ${\hat\eta}_{cc}$ (28\%),
${\hat B}_K$ (16\%), $|V_{ub}/V_{cb}|$ (15\%) and
$\fbd\sqrt{\hat{B}_{B_d}}$ (19\%). Of these, the latter three are
extremely important in defining the allowed $\rho$--$\eta$ region (the
large error on ${\hat\eta}_{cc}$ does not affect the fit very much).
The errors on two of these quantities --- ${\hat B}_K$ and
$\fbd\sqrt{\hat{B}_{B_d}}$ --- are purely theoretical in origin, and
the error on $|V_{ub}/V_{cb}|$ has a significant theoretical component
(model dependence). Thus, the present uncertainty in the shape of the
unitarity triangle is due in large part to theoretical errors.
Reducing these errors will be quite important in getting a precise
profile of the unitarity triangle and the CP-violating phases.

\begin{table}
\hfil
\vbox{\offinterlineskip
\halign{&\vrule#&
 \strut\quad#\hfil\quad\cr
\noalign{\hrule}
height2pt&\omit&&\omit&\cr
& Parameter && Value & \cr
height2pt&\omit&&\omit&\cr
\noalign{\hrule}
height2pt&\omit&&\omit&\cr
& $\lambda$ && $0.2196$ & \cr
& $\vert V_{cb} \vert $ && $0.0395 \pm 0.0017$ & \cr
& $\vert V_{ub} / V_{cb} \vert$ && $0.093 \pm 0.014$ & \cr
& $\abseps$ && $(2.280 \pm 0.013) \times 10^{-3}$ & \cr
& $\Delta M_d$ && $(0.471 \pm 0.016)~(ps)^{-1}$ & \cr
& $\Delta M_s$ && $ > 12.4 ~(ps)^{-1}$ & \cr 
& $\overline{\mt}(\mt(pole))$ && $(165 \pm 5)$ GeV & \cr
& $\overline{\mc}(\mc(pole))$ && $1.25 \pm 0.05$ GeV & \cr
& $\hat{\eta}_B$ && $0.55$ & \cr
& $\hat{\eta}_{cc} $ && $1.38 \pm 0.53$ & \cr
& $\hat{\eta}_{ct} $ && $0.47 \pm 0.04$ & \cr
& $\hat{\eta}_{tt} $ && $0.57$ & \cr
& $\hat{B}_K$ && $0.94 \pm 0.15$ & \cr
& $\fbd\sqrt{\hat{B}_{B_d}} $ && $215 \pm 40$ MeV & \cr
& $\xi_s $ && $1.14 \pm 0.06$  & \cr
height2pt&\omit&&\omit&\cr
\noalign{\hrule}}}
\caption{Data used in the CKM fits. Values of the hadronic quantities
  $\hat{B}_K$, $\fbd \protect\sqrt{\hat{B}_{B_d}}$, and $\xi_s$ are
  taken from the lattice QCD results \protect\cite{Draper98,Sharpe98}.
  The remaining theoretical numbers are discussed in the text. The
  value for $\Delta M_d$ and the $95\%$ C.L.\ lower limit on $\Delta
  M_s$ are taken from the LEP Electroweak group
  \protect\cite{Parodiconf98}. All other experimental numbers are
  taken from the Particle Data Group \protect\cite{PDG98}.}
\label{datatable}
\end{table}

There are two other measurements which should be mentioned here.

First, the KTEV collaboration \cite{KTEV99} has recently reported a
measurement of direct CP violation in the $K$ sector through the ratio
$\epsilon^\prime/\epsilon$, with
\beq
{\rm Re} (\epsilon^\prime/\epsilon) = \left( 28.0 \pm 3.0 (\mbox{stat}) 
\pm 2.6 (\mbox{syst}) \pm 1.0(\mbox{MC stat})\right) \times 10^{-4} ~,
\eeq
in agreement with the earlier measurement by the CERN experiment NA31
\cite{NA31}, which reported a value of $(23 \pm 6.5) \times 10^{-4}$
for the same quantity. The present world average is ${\rm Re}
(\epsilon^\prime/\epsilon) =(21.8 \pm 3.0) \times 10^{-4}$. This
combined result excludes the superweak model \cite{superweak} by more
than $7\sigma$. 

A great deal of theoretical effort has gone into calculating this
quantity at next-to-leading order accuracy in the SM
\cite{Buraseps,Martinellieps,Bertolinieps}. The result of this
calculation has been summarized in a succint form by Buras and
Silvestrini \cite{BS98}:
\beq
{\rm Re} (\epsilon^\prime/\epsilon) = {\rm Im} \lambda_t 
\left[ -1.35 + R_s \left( 1.1 \vert r_Z^{(8)} \vert B_6^{(1/2)} 
+ (1.0 -0.67 \vert r_Z^{(8)} \vert) B_8^{(3/2)} \right) \right]~.
\label{epsilonp}
\eeq
Here $\lambda_t= V_{td}V_{ts}^* = A^2 \lambda^4 \eta$ and $r_Z^{(8)}$
represents the short-distance contribution, which at the NLO precision
is estimated to lie in the range $ 6.5 \leq \vert r_Z^{(8)} \vert \leq
8.5$ \cite{Buraseps,Martinellieps}. The quantities
$B_6^{(1/2)}=B_6^{(1/2)}(m_c)$ and $B_8^{(3/2)}=B_8^{(3/2)}(m_c)$ are
the matrix elements of the $\Delta I=1/2$ and $\Delta I =3/2$
operators $O_6$ and $O_8$, respectively, calculated at the scale
$\mu=m_c$. Lattice-QCD \cite{B68-Latt} and the $1/N_c$ expansion
\cite{BurasNc} yield:
\beq
0.8 \leq B_6^{(1/2)} \leq 1.3,~~~~~~~~~~~0.6 \leq B_8^{(3/2)} \leq 1.0~.
\eeq
Finally, the quantity $R_s$ in Eq.~(\ref{epsilonp}) is defined as:
\beq
R_s \equiv \left( \frac{150~\mbox{MeV}}{m_s(m_c) + m_d(m_c)} \right)^2 ~,
\label{Rsdef}
\eeq
essentially reflecting the $s$-quark mass dependence. The present
uncertainty on the CKM matrix element is $\pm 23\%$, which is already
substantial. However, the theoretical uncertainties related to the
other quantities discussed above are considerably larger. For
example, the ranges $\epsilon^\prime/\epsilon=(5.3 \pm 3.8) \times
10^{-4}$ and $\epsilon^\prime/\epsilon=(8.5 \pm 5.9) \times 10^{-4}$,
assuming $m_s(m_c)=150\pm 20$ MeV and $m_s(m_c)=125\pm 20$ MeV,
respectively, have been quoted as the best representation of the
status of $\epsilon^\prime/\epsilon$ in the SM \cite{Burasreview98}.
These estimates are somewhat on the lower side compared to the data
but not inconsistent. (For some recent speculations on new physics,
see Refs.~\cite{KNS99,MM99}.)

Thus, whereas $\epsilon^\prime/\epsilon$ represents a landmark
measurement, removing the superweak model of Wolfenstein and its kith
and kin from further consideration, its impact on CKM phenomenology,
particularly in constraining the CKM parameters, is marginal.
Probably the best use of this measurement is to constrain the
$s$-quark mass, which at present has considerable uncertainty. For
this reason we do not include the measurement of
$\epsilon^\prime/\epsilon$ in the CKM fits presented here.

Second, the CDF collaboration has recently made a measurement of $\sin
2\beta$ \cite{CDF99}. In the Wolfenstein parametrization, $-\beta$ is
the phase of the CKM matrix element $V_{td}$. From Eq.~(\ref{CKM}) one
can readily find that
\beq
\sin (2 \beta) = \frac{2\eta(1-\rho)}{(1-\rho)^2 + \eta^2} ~.
\eeq
Thus, a measurement of $\sin 2\beta$ would put a strong contraint on
the parameters $\rho$ and $\eta$. However, the CDF measurement gives
\cite{CDF99}
\beq
\sin 2\beta = 0.79^{+0.41}_{-0.44} ~,
\eeq
or $\sin 2\beta > 0$ at 93\% C.L. As we will see in the next section,
this constraint is quite weak -- the other measurements already
constrain $0.52 \le \sin 2\beta \le 0.94$ at the 95\% C.L.\ in the SM.
(The CKM fits reported in Refs.~\cite{Parodiconf98,Mele98,PRS98} yield
similar ranges.)  In light of this, and given that it is not clear how
to combine the above measurement (which allows for unphysical values
of $\sin 2\beta$) with the other data, we have not included this
measurement in our fits.

\subsection{SM Fits}

In order to find the allowed region in $\rho$--$\eta$ space, i.e.\ the
allowed shapes of the unitarity triangle, the computer program MINUIT
is used to fit the parameters to the constraints described above. In
the fit, we allow ten parameters to vary: $\rho$, $\eta$, $A$, $m_t$,
$m_c$, $\eta_{cc}$, $\eta_{ct}$, $f_{B_d} \sqrt{\hat{B}_{B_d}}$,
$\hat{B}_K$, and $\xi_s$. The $\Delta M_s$ constraint is included using
the amplitude method. 

The allowed (95\% C.L.) $\rho$--$\eta$ region is shown in
Fig.~\ref{rhoeta1}. The best fit has $(\rho,\eta) = (0.20,0.37)$.

\begin{figure}
\vskip -1.0truein
\centerline{\epsfxsize 3.5 truein \epsfbox {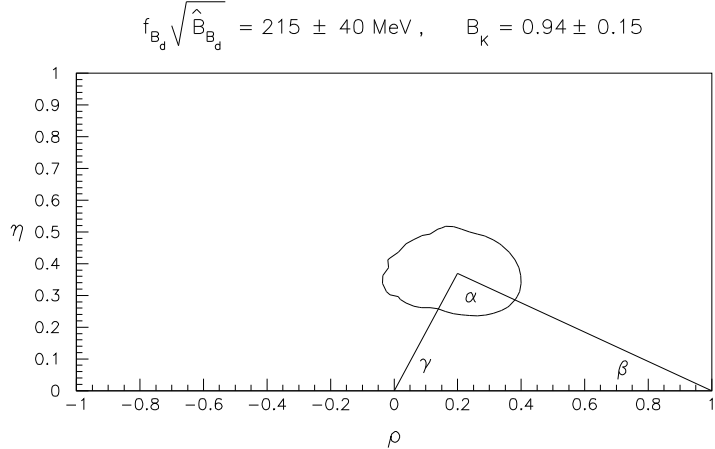}}
\vskip -1.4truein
\caption{Allowed region in $\rho$--$\eta$ space in the SM, from a
  fit to the ten parameters discussed in the text and given in Table
  \protect{\ref{datatable}}. The limit on $\Delta M_s$ is included
  using the amplitde method of Ref.~\protect\cite{Moser97}. The
  theoretical errors on $\fbd\protect\sqrt{\hat{B}_{B_d}}$,
  $\hat{B}_K$ and $\xi_s$ are treated as Gaussian. The solid line
  represents the region with $\chi^2=\chi_{min}^2+6$ corresponding to
  the 95\% C.L.\ region. The triangle shows the best fit.}
\label{rhoeta1}
\end{figure}

There is an alternative way to include the $\Delta M_s$ constraint,
one which we have used in the past \cite{AL96}. In this case the
constraint is excluded from the fit, but rather one cuts away a region
of $\rho$--$\eta$ space by superimposing a line corresponding to a
particular value of $\xi_s$. Using the 95\% C.L.\ limit on $\Delta M_s$
and allowing $\Delta M_d$ to fluctuate upward by $1\sigma$ from its
central value, one obtains $\Delta M_s/\Delta M_d > 25.5$. This yields
the following constraint on $\rho$ and $\eta$:
\beq
\lambda^2 \left[ (1-\rho)^2 + \eta^2 \right] = 
\left( {\Delta M_d \over \Delta M_s} \right) \xi_s^2
\le {1\over 25.5} \left( 1.14 \pm 0.06 \right)^2 ~.
\eeq
We take three candidate values for $\xi_s$: 1.08, 1.14, 1.20. The
results are shown in Fig.~\ref{xsrhoeta1}. The best fit has
$(\rho,\eta) = (0.19,0.37)$.

\begin{figure}
\vskip -1.0truein
\centerline{\epsfxsize 3.5 truein \epsfbox {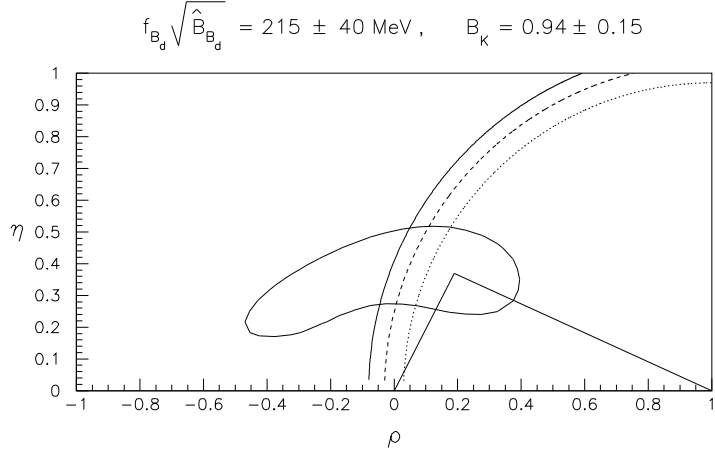}}
\vskip -1.4truein
\caption{Allowed region in $\rho$--$\eta$ space in the SM. The 
  $\Delta M_s$ limit is not included in the fit, but rather imposed by
  cutting away a region of the space. The disallowed region is shown
  for three choices of the parameter $\xi_s$: 1.08 (dotted line), 1.14
  (dashed line), 1.20 (solid line). In all cases, the region to the
  left of the curve is ruled out.}
\label{xsrhoeta1}
\end{figure}

A comparison of Figs.~\ref{rhoeta1} and \ref{xsrhoeta1} reveals that
the allowed regions are similar, though not identical, for the two
methods of including the $\Delta M_s$ constraint. Henceforth, all our
results will be presented using the fit which includes this constraint
via the amplitude method.

The CP angles $\alpha$, $\beta$ and $\gamma$ can be measured in
CP-violating rate asymmetries in $B$ decays \cite{BCPasym}. For
example, the asymmetries in $\bd(t) \to \pi^+\pi^-$ and $\bd(t) \to
J/\Psi \ks$ probe $\sin 2\alpha$ and $\sin 2\beta$, respectively. The
angle $\gamma$ can be extracted from $B^\pm \to D K^\pm$ \cite{BtoDK}
or $\bs(t) \to D_s^\pm K^\mp$ \cite{ADK}. The function in this case is
$\sin^2 \gamma$.

{}From Fig.~\ref{triangle}, it is clear that these angles can be
expressed in terms of $\rho$ and $\eta$. Thus, different shapes of the
unitarity triangle are equivalent to different values of the CP
angles. Referring to Fig.~\ref{rhoeta1}, we note that the preferred
(central) values of these angles are $(\alpha,\beta,\gamma) =
(93^\circ,25^\circ,62^\circ)$. The allowed ranges at 95\% C.L. are
\begin{eqnarray}
\label{CPangleregion}
65^\circ \le & \alpha & \le 123^\circ \nn\cr
16^\circ \le & \beta & \le 35^\circ \nn\cr
36^\circ \le & \gamma & \le 97^\circ
\end{eqnarray}
or, equivalently,
\begin{eqnarray}
-0.91 \le & \sin 2\alpha & \le 0.77 \nn\cr
0.52  \le & \sin 2\beta & \le 0.94  \nn\cr
0.35  \le & \sin^2 \gamma & \le 1.00 ~.
\end{eqnarray}

Of course, the values of $\alpha$, $\beta$ and $\gamma$ are
correlated, i.e.\ they are not all allowed simultaneously. After all,
the sum of these angles must equal $180^\circ$. We illustrate these
correlations in Figs.~\ref{alphabetacorr} and \ref{alphagammacorr}.
The plots with $f=0$ correspond to the SM. (Nonzero values of $f$
correspond to supersymmetric models, which will be discussed in detail
in the next section.) Fig.~\ref{alphabetacorr} shows the allowed
region in $\sin 2\alpha$--$\sin 2\beta$ space allowed by the data. And
Fig.~\ref{alphagammacorr} shows the allowed (correlated) values of the
CP angles $\alpha$ and $\gamma$. This correlation is roughly linear,
due to the relatively small allowed range of $\beta$
(Eq.~(\ref{CPangleregion})).

\begin{figure}
\vskip -2.0truein
\centerline{\epsfxsize 8.0 truein \epsfbox {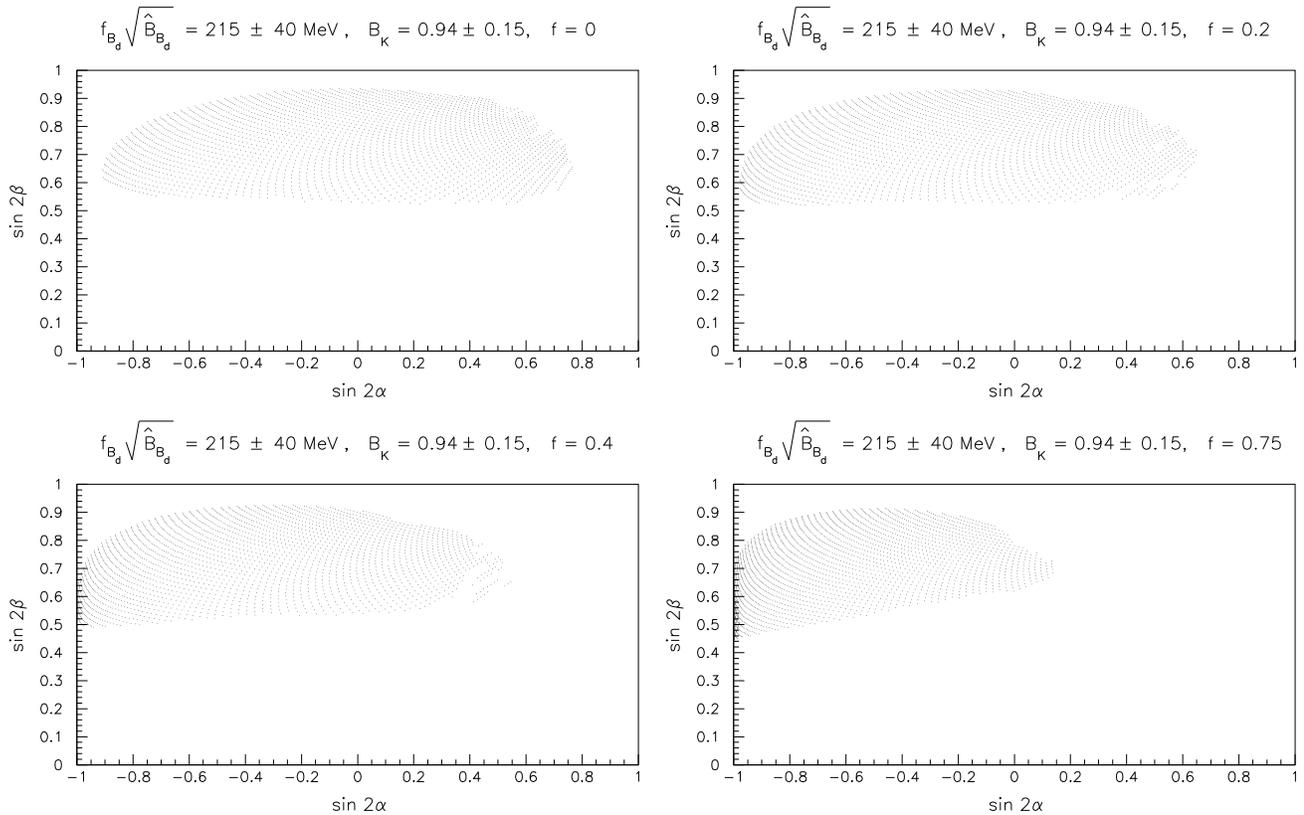}}
\vskip -4.7truein
\caption{Allowed 95\% C.L. region of the CP-violating quantities 
  $\sin 2\alpha$ and $\sin 2\beta$, from a fit to the data given in
  Table \protect{\ref{datatable}}. The upper left plot ($f=0$)
  corresponds to the SM, while the other plots ($f=0.2$, 0.4, 0.75)
  correspond to various SUSY models.}
\label{alphabetacorr}
\end{figure}

\begin{figure}
\vskip -2.0truein
\centerline{\epsfxsize 8.0 truein \epsfbox {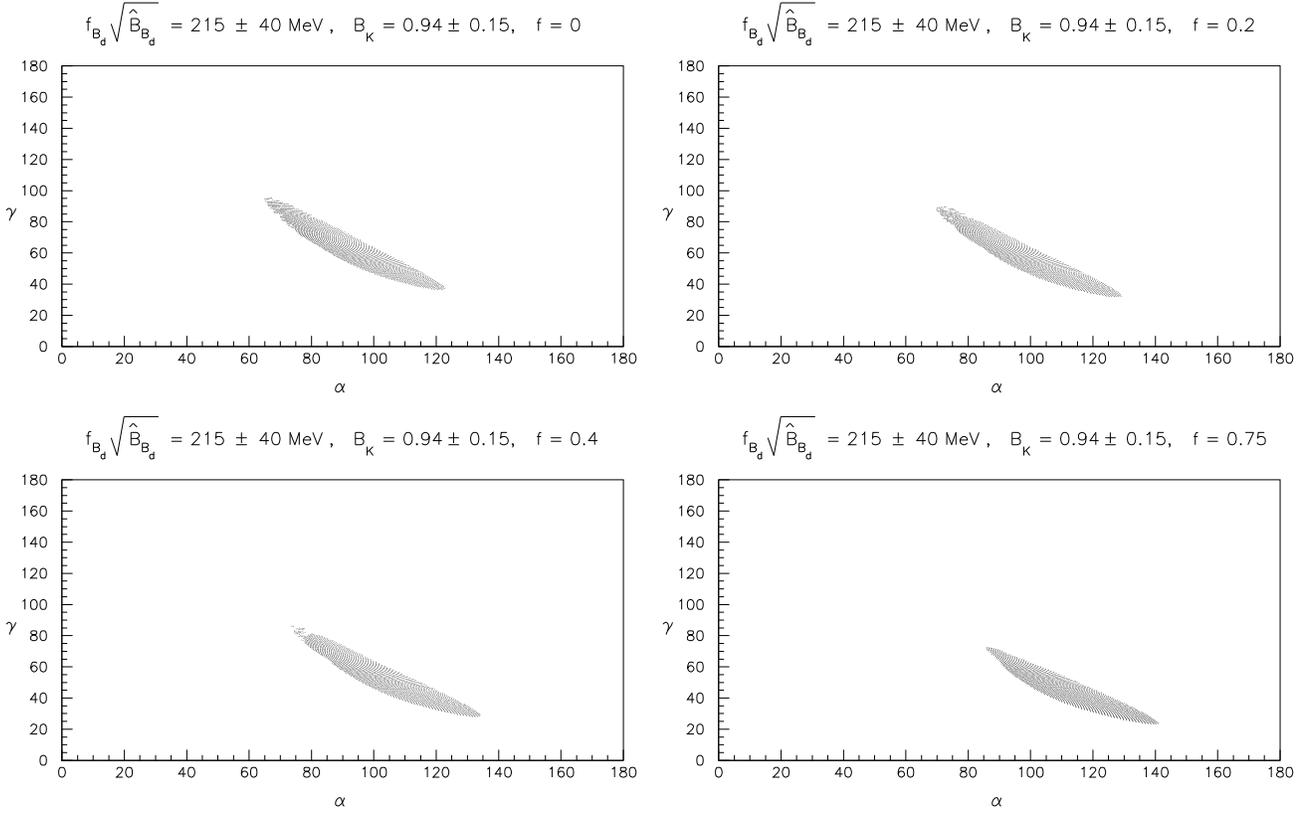}}
\vskip -4.7truein
\caption{Allowed 95\% C.L. region of the CP-violating quantities 
  $\alpha$ and $\gamma$, from a fit to the data given in Table
  \protect{\ref{datatable}}. The upper left plot ($f=0$) corresponds
  to the SM, while the other plots ($f=0.2$, 0.4, 0.75) correspond to
  various SUSY models.}
\label{alphagammacorr}
\end{figure}

\section{Unitarity Triangle: A SUSY Profile}

In this section we examine the profile of the unitarity triangle in
supersymmetric (SUSY) theories. As we will see below, the most general
models contain a number of unconstrained phases and so are not
sufficiently predictive to perform such an analysis. However, there is
a class of SUSY models in which these phases are constrained to be
approximately zero, which greatly increases the predictivity. Not
surprisingly, the calculations of the SUSY contributions to measured
quantities are also more advanced in these models, having been
performed at the next-to-leading-order (NLO) level. In the following
subsections, we discuss aspects of more general SUSY theories, as well
as the details of that class of theories whose effects on the
unitarity triangle can be directly analyzed.

\subsection{Flavour Violation in SUSY Models - Overview}

We begin with a brief review of flavour violation in the minimal
supersymmetric standard model (MSSM).

The low energy effective theory in the MSSM can be specified in terms
of the chiral superfields for the three generations of quarks ($Q_i$,
$U_i^c$, and $D_i^c$) and leptons ($L_i$ and $E_i^c$), chiral
superfields for two Higgs doublets ($H_1$ and $H_2$), and vector
superfields for the gauge group $SU(3)_C \times SU(2)_I \times U(1)_Y$
\cite{Nilles84}. The superpotential is given by
\beq
W_{\sss MSSM} = f_D^{ij} Q_i D_j H_1 + f_U^{ij} Q_i U_j H_2 +
                  f_L^{ij} E_i L_j H_1 + \mu H_1 H_2.
\label{superpotential}
\eeq
%
The indices $i,j=1,2,3$ are generation indices and $f_D^{ij}$,
$f_U^{ij}$, $f_L^{ij}$ are Yukawa coupling matrices in the generation
space. A general form of the soft SUSY-breaking term is given by
\begin{eqnarray}
-{\cal L}_{\mbox{soft}} &=& (m_Q^2)^i_j \tilde{q}_i \tilde{q}^{\dagger j}
+ (m_D^2)^i_j \tilde{d}_i \tilde{d}^{\dagger j}
+ (m_U^2)^i_j \tilde{u}_i \tilde{u}^{\dagger j}
+ (m_E^2)^i_j \tilde{e}_i \tilde{e}^{\dagger j}
+ (m_L^2)^i_j \tilde{\ell}_i \tilde{\ell}^{\dagger j} \nonumber\\
&+& \Delta_1^2 h_1^\dagger h_1 + \Delta_2^2 h_2^\dagger h_2
-(B\mu h_1 h_2 + h.c.) \nonumber\\
&+&(A_D^{ij} \tilde{q}_i \tilde{d}_j h_1 + A_U^{ij} \tilde{q}_i \tilde{u}_j 
h_2 + A_L^{ij} \tilde{e}_i \tilde{\ell}_j h_1 + h.c.) \nonumber\\
&+& (\frac{M_1}{2}\tilde{B}\tilde{B} + \frac{M_2}{2}\tilde{W}\tilde{W}
+\frac{M_3}{2}\tilde{G}\tilde{G} + h.c.),
\label{softsusy}
\end{eqnarray}
where $\tilde{q}_i$, $\tilde{u}_i$, $\tilde{d}_i$, $\tilde{\ell}_i$,
$\tilde{e}_i$, $h_1$ and $h_2$ are scalar components of the superfields
$Q_i$, $U_i$, $D_i$, $L_i$, $E_i$, $H_1$ and $H_2$, respectively, and
$\tilde{B}$, $\tilde{W}$ and $\tilde{G}$ are the $U(1)$, $SU(2)$ and $SU(3)$
gauge fermions, respectively.

In fact, there are many variants of the MSSM. All have the same
particle content, but the mass hierarchies and supersymmetry-breaking
parameters are different. As a result, they lead to different
low-energy predictions.

For example, in supergravity (SUGRA) models, the SUSY-breaking
parameters in the MSSM are assumed to have a simple structure at the
Planck scale \cite{Nilles84}. Neglecting the difference between the
Planck and the GUT scales, and defining this scale by $X$, one can
express them as follows:
\begin{eqnarray}
(m_Q^2)^i_j &=& (m_E^2)^i_j = m_0^2 \delta^i_j, \nonumber\\
(m_D^2)^i_j &=& (m_U^2)^i_j =(m_L^2)^i_j = m_0^2 \delta^i_j, \nonumber\\ 
\Delta_1^2 &=& \Delta_2^2 = \Delta_0^2, \nonumber\\
A_D^{ij} &=& f_{DX}^{ij} A_X m_0, ~~ A_L^{ij} = f_{LX}^{ij} A_X m_0, ~~
A_U^{ij} = f_{UX}^{ij} A_X m_0, \nonumber\\
M_1 &=& M_2 = M_3 =M_{gX} ~.
\label{sugraconst}
\end{eqnarray}
In the minimal SUGRA case, $m_0=\Delta_0$, while in the non-minimal
SUGRA case, one takes $m_0$ and $\Delta_0$ as independent parameters.

In general, MSSM models have three physical phases, apart from the QCD
vacuum parameter $\bar{\theta}_{\sss QCD}$ which we shall take to be
zero. The three phases are: (i) the CKM phase represented here by the
Wolfenstein parameter $\eta$, (ii) the phase $\theta_A=\arg (A)$, and
(iii) the phase $\theta_\mu=\arg (\mu)$ \cite{DGH85}. The last two
phases are peculiar to SUSY models and their effects must be taken
into account in a general supersymmetric framework. In particular, the
CP-violating asymmetries which result from the interference between
mixing and decay amplitudes can
produce non-standard effects. Concentrating here on the $\Delta B=2$
amplitudes, two new phases $\theta_d$ and $\theta_s$ arise, which can
be parametrized as follows \cite{effsusy}:
\beq
\theta_{d,s} = \frac{1}{2} \arg \left(\frac{\langle B_{d,s} \vert {\cal 
H}_{eff}^{\sss SUSY} \vert \bar{B}_{d,s} \rangle}{\langle B_{d,s} \vert 
{\cal H}_{eff}^{\sss SM} \vert \bar{B}_{d,s} \rangle} \right) ~,
\eeq
where ${\cal H}^{\sss SUSY}$ is the effective Hamiltonian including
both the SM degrees of freedom and the SUSY contributions. Thus,
CP-violating asymmetries in $B$ decays would involve not only the
phases $\alpha$, $\beta$ and $\gamma$, defined previously, but
additionally $\theta_d$ or $\theta_s$. In other words, the SUSY
contributions to the real parts of $M_{12}(B_d)$ and $M_{12}(B_s)$ are
{\it no longer proportional} to the CKM matrix elements $V_{td}
V_{tb}^*$ and $V_{ts}V_{tb}^*$, respectively. If $\theta_d$ or
$\theta_s$ were unconstrained, one could not make firm predictions
about the CP asymmetries in SUSY models. In this case, an analysis of
the profile of the unitarity triangle in such models would be futile.

However, the experimental upper limits on the electric dipole moments
(EDM) of the neutron and electron \cite{PDG98} do provide constraints
on the phases $\theta_\mu$ and $\theta_A$ \cite{FOS95}. In SUGRA
models with {\it a priori} complex parameters $A$ and $\mu$, the phase
$\theta_\mu$ is strongly bounded with $\theta_\mu < 0.01 \pi$
\cite{Nihei97}. The phase $\theta_A$ can be of $O(1)$ in the small
$\theta_\mu$ region, as far as the EDMs are concerned. In both the
$\Delta S=2$ and $\Delta B=2$ transitions, and for low-to-moderate
values of $\tan \upsilon$ \footnote{In supersymmetric jargon, the
  quantity $\tan \beta$ is used to define the ratio of the two vacuum
  expectation values (vevs) $\tan \beta \equiv v_u/v_d$, where
  $v_d(v_u)$ is the vev of the Higgs field which couples exclusively
  to down-type (up-type) quarks and leptons. (See, for example, the
  review by Haber in Ref.~\cite{PDG98}). However, in discussing
  flavour physics, the symbol $\beta$ is traditionally reserved for
  one of the angles of the unitarity triangle. To avoid confusion, we
  will call the ratio of the vevs $\tan \upsilon$.}, it has been shown
that $\theta_A$ does not change the phase of either the matrix element
$M_{12}(K)$ \cite{DGH85} or of $M_{12}(B)$ \cite{Nihei97}. Hence, in
SUGRA models, $\arg M_{12}(B)|_{\sss SUGRA} = \arg M_{12}(B)|_{\sss
  SM}=\arg (\xi_t^2)$, where $\xi_t=V_{td}^*V_{tb}$. Likewise, the
phase of the SUSY contribution in $M_{12}(K)$ is aligned with the
phase of the $t\bar{t}$-contribution in $M_{12}(K)$, given by $\arg
(V_{td}V_{ts}^*)$. Thus, in SUGRA models, one can effectively set
$\theta_d \simeq 0$ and $\theta_s \simeq 0$, so that the CP-violating
asymmetries give information about the SM phases $\alpha$, $\beta$ and
$\gamma$. Hence, an analysis of the UT and CP-violating phases
$\alpha$, $\beta$ and $\gamma$ can be carried out in a very similar
fashion as in the SM, taking into account the additional contributions
to $M_{12}(K)$ and $M_{12}(B)$.

For large-$\tan\upsilon$ solutions, one has to extend the basis of
$H_{eff} (\Delta B=2)$ so as to include new operators whose contribution 
is small in the low-$\tan\upsilon$ limit. The resulting effective Hamiltonian
is given by
\beq 
H_{eff}(\Delta B=2) = \frac{G_F^2M_W^2}{2\pi^2} \sum_{i=1}^{3} 
C_i(\mu)O_{i}~,
\label{SMeffham}
\eeq
where $O_1=\bar{d}_L^\alpha\gamma_\mu b_L^\alpha \bar{d}_L^\beta
\gamma^\mu b_L^\beta$, $O_2= \bar{d}_L^\alpha b_R^\alpha
\bar{d}_L^\beta b_R^\beta$ and $O_3= \bar{d}_L^\alpha b_R^\beta
\bar{d}_L^\beta b_R^\alpha$ and $C_i$ are the Wilson coefficients
\cite{Brancoetal,CS98}. The coefficients $C_1(\mu)$ and $C_2(\mu)$ are
real relative to the SM contribution. However, the chargino
contributions to $C_3(\mu)$ are generally complex relative to the SM
contribution and can generate a new phase shift in the
$B^0$--$\overline{B^0}$ mixing amplitude \cite{DMV98,BK98-1}. This
effect is in fact significant for large $\tan \upsilon$
\cite{Brancoetal}, since $C_3(\mu)$ is proportional to $(m_b/m_W\cos
\beta)^2$. How large this additional phase $(\theta_d $ and
$\theta_s$) can be depends on how the constraints from EDM are
imposed. For example, Baek and Ko \cite{BK98-1} find that in the MSSM
without imposing the EDM constraint, one has $2 \vert \theta_d \vert
\leq 6^\circ$ for a light stop and large $\tan \upsilon$ but this phase
becomes practically zero if the EDM constraints \cite{CKP98} are
imposed. This aspect of the analysis in Ref.~\cite{BK98-1}, though
done without invoking the SUGRA model mass relations, should also hold
in SUGRA models.


In a more general supersymmetric scenario, the SUSY effects may lead
to additional phases and the constraints from EDMs may not be
sufficient to effectively bound them. (A recent example is the
so-called ``effective supersymmetry'' \cite{effsusy}.) As discussed
above, such models do not make clean predictions for the CP
asymmetries, so that one cannot analyse the profile of the unitarity
triangle in this case. Instead, flavour-changing effects in $K$ and
$B$ decays in these scenarios will have to be disentangled along the
lines suggested in Refs.~\cite{effsusy,SW96,Randallsu}.
These theories may give rise to flavour-violation effects which are
uncharacteristic of the SM and MSSM. In particular,
they may lead to a measurable charge asymmetry in
semileptonic decays of the $B^0$ and $\overline{B^0}$
\cite{Randallsu}, which is estimated to be negligible in both the SM
and MSSM/SUGRA models discussed above. However, some technical
aspects in the most general supersymmetric theories remain to be
worked out. For example, the complete NLO corrections for $\delmd$,
$\delms$ and $\abseps$ have not been calculated in these theories.
The bag constants which correct for the vacuum insertion approximation
in the matrix elements of the operators in $H_{eff}^{\Delta S=2}$ have
been calculated using lattice-QCD techniques \cite{Scimemi98}, but the
corresponding estimates for the bag parameters $H_{eff}^{\Delta B=2}$
are not yet available. Likewise, the NLO calculations in the general
supersymmetric case for the decay $B \to X_s \gamma$ are not yet at
hand. This last ingredient enters the estimates of the quantities
$\delmd$, $\delms$ and $\abseps$ indirectly as the measured branching
ratio ${\cal B}(B \to X_s \gamma)$ provides rather stringent
constraints on the allowed supersymmetric parameters.

In view of the foregoing, we shall restrict ourselves to a class of
SUSY models in which the following features, related to flavour
mixing, hold:
\begin{itemize}

\item The squark flavour mixing matrix which diagonalizes the squark
mass matrix is approximately the same as the corresponding quark
mixing matrix $V_{CKM}$, apart from the left-right mixing of the top
squarks.

\item The first- and second-generation squarks with the same gauge
quantum numbers remain highly degenerate in masses but the
third-generation squarks, especially the top squark, can be
significantly lighter due to the renormalization effect of the top
Yukawa coupling constants.

\item The phases $\theta_d$ and $\theta_s$ are negligible in the entire 
$\tan \upsilon$ plane, once the constraints from the EDMs of neutron and 
lepton are consistently  imposed.

\end{itemize}
These features lead to an enormous simplification in the flavour
structure of the SUSY contributions to flavour-changing processes. In
particular, SUSY contributions to the transitions $b \to s$, $b \to d$
and $s \to d$ are proportional to the CKM factors, $V_{tb}V_{ts}^*$,
$V_{tb}V_{td}^*$ and $V_{ts}V_{td}^*$, respectively. Similarly, the
SUSY contributions to the mass differences $M_{12}(B_s)$,  
$M_{12}(B_d)$ and $M_{12}(K)$ are proportional to the CKM factors
$(V_{tb}V_{ts}^*)^2$, $(V_{tb}V_{td}^*)^2$ and $(V_{ts}V_{td}^*)^2$,
respectively. These are precisely the same factors which govern the
contribution of the top quark in these transitions in the standard
model. Thus, the supersymmetric contributions can be implemented in a
straightforward way by adding a (supersymmetric) piece in each of the
above mentioned amplitudes to the corresponding top quark contribution
in the SM.

\subsection{NLO Corrections to $\delmd$, $\delms$ and $\epsilon$ in Minimal 
SUSY Flavour Violation}

A number of SUSY models share the features mentioned in the previous
subsection,
and the supersymmetric contributions to the mass differences
$M_{12}(B)$ and $M_{12}(K)$ have been analyzed in a number of papers
\cite{Nihei97,Brancoetal,Gotoetal96,Gotoetal97,Gotoetal98-1,Gotoetal98-2},
following the pioneering work of Ref.~\cite{BBMR91}. Following these
papers, $\delmd$ can be expressed as:
\beq
\delmd =  \frac{G_F^2}{6\pi^2}M_W^2M_B\left(\fbb\right)\hat{\eta}_B
\left[A_{SM}(B) + A_{H^\pm}(B) + A_{\chi^\pm}(B) + A_{\tilde{g}}(B) \right]~,
\label{delmdsusy}
\eeq
where the function $A_{SM}(B)$ can be written by inspection from
Eq.~(\ref{bdmixing}):
\beq
A_{SM}(B)= y_t f_2(y_t) \vert V_{td}^* V_{tb} \vert^2 ~.
\eeq
The expressions for $A_{H^\pm}(B)$, $A_{\chi^\pm}(B)$ and
$A_{\tilde{g}}(B)$ are obtained from the SUSY box diagrams.
Here, $H^\pm$, $\chi^\pm_j$, $\tilde{t}_a$ and $\tilde{d}_i$ represent,
respectively, the charged Higgs, chargino, stop and down-type
squarks. The contribution of the intermediate states involving
neutralinos is small and usually neglected. The expressions for
$A_{H^\pm}(B)$, $A_{\chi^\pm}(B)$ and $A_{\tilde{g}}(B)$ are given
explicitly in the literature \cite{Brancoetal,Gotoetal96,BBMR91}.

We shall not be using the measured value of the mass difference $\Delta
M_K$ due to the uncertain contribution of the long-distance
contribution. However, $\abseps$ is a short-distance dominated
quantity and in supersymmetric theories can be expressed as follows:
\beq
\abseps = \frac{G_F^2f_K^2M_KM_W^2}{6\sqrt{2}\pi^2\Delta M_K}
\hat{B}_K
\left[\mbox{Im}~A_{SM}(K) + \mbox{Im}~A_{H^\pm}(K) + 
\mbox{Im}~A_{\chi^\pm}(K) + \mbox{Im}~A_{\tilde{g}}(K) \right]~,
\label{epsilonsusy}
\eeq
where, again by inspection with the SM expression for $\abseps$ given 
in Eq.~(\ref{eps}), one has   
\beq
\mbox{Im}~A_{SM}(K) =
A^2\lambda^6\eta
\bigl(y_c\left\{\hat{\eta}_{ct}f_3(y_c,y_t)-\hat{\eta}_{cc}\right\}
+~\hat{\eta}_{tt}y_tf_2(y_t)A^2\lambda^4(1-\rho)\bigr)
\label{asm}
\eeq
The expressions for $\mbox{Im}~A_{H^\pm}(K)$,
$\mbox{Im}~A_{\chi^\pm}(B)$ and $\mbox{Im}~A_{\tilde{g}}(B)$ can be
found in Refs.~\cite{Brancoetal,Gotoetal96,BBMR91}.

For the analysis reported here, we follow the scenario called {\it
  minimal flavour violation} in Ref.~\cite{CDGG98}. In this class of
supersymmetric theories, apart from the SM degrees of freedom, only
charged Higgses, charginos and a light stop (assumed to be
right-handed) contribute, with all other supersymmetric particles
integrated out. This scenario is effectively implemented in a class of
SUGRA models (both minimal and non-minimal) and gauge-mediated models
\cite{Dine93}, in which the first two squark generations are heavy and
the contribution from the intermediate gluino-squark states is small
\cite{Nihei97,Gotoetal96,Gotoetal97,Gotoetal98-1,Gotoetal98-2}.

For these models, the next-to-leading-order (NLO) corrections for
$\delmd$, $\delms$ and $\abseps$ can be found in Ref.~\cite{KS98}.
Moreover, the branching ratio ${\cal B}(B \to X_s \gamma)$ has been
calculated in Ref.~\cite{CDGG98}. We make use of this information and
quantitatively examine the unitarity triangle, CP-violating
asymmetries and their correlations for this class of supersymmetric
theories. The phenomenological profiles of the unitarity triangle and
CP phases for the SM and this class of supersymmetric models can thus
be meaningfully compared. Given the high precision on the phases
$\alpha$, $\beta$ and $\gamma$ expected from experiments at
$B$-factories and hadron colliders, a quantitative comparison of this
kind could provide a means of discriminating between the SM and this
class of MSSM's.

The NLO QCD-corrected effective Hamiltonian
for $\Delta B=2$ transitions in the minimal flavour violation SUSY
framework can be expressed as follows \cite{KS98}:
\beq
H_{eff} = \frac{G_F^2}{4 \pi^2} (V_{td}V_{tb}^*)^2 \hat{\eta}_{2,S}(B) S 
O_{LL}~,
\label{effHsusy}
\eeq 
where the NLO QCD correction factor $\hat{\eta}_{2,S}(B)$ is given by
\cite{KS98}:
\beq
\hat{\eta}_{2,S}(B) = \alpha_s(m_W) \gamma^{(0)}/(2 \beta_{n_f}^{(0)}) \left[
1 + \frac{\alpha_s(m_W)}{4 \pi} \left(\frac{D}{S} + Z_{n_f}\right) \right]~,
\label{eta2s}
\eeq
in which $n_f$ is the number of active quark flavours (here $n_f=5$),
and the quantities $Z_{n_f}$, $\gamma^{(0)}$ and $\beta_{n_f}^{(0)}$
are defined below. The operator $O_{LL}=O_1$ is the one which is
present in the SM, previously defined in the discussion following
Eq.~(\ref{SMeffham}).
The explicit expression for the function $S$ can be obtained from
Ref.~\cite{BBMR91} and for $D$ it is given in Ref.~\cite{KS98}, where
it is derived in the NDR (naive dimensional regularization) scheme
using $\overline{MS}$-renormalization.

The Hamiltonian given above for $B_d^0$--$\overline{B_d^0}$ mixing
leads to the mass difference
\beq
\delmd = \frac{G_F^2}{6 \pi^2} (V_{td}V_{tb}^*)^2 \hat{\eta}_{2,S}(B) S
(\fbd^2 \hat{B}_{B_d})~.
\label{deltamdsusy}
\eeq
The corresponding expression for $\delms$ is obtained by making the
appropriate replacements. Since the QCD correction factors are
identical for $\delmd$ and $\delms$, it follows that the quantities
$\delmd$ and $\delms$ are enhanced by the same factor in minimal
flavour violation supersymmetry, as compared to their SM values, but
the ratio $\delms/\delmd$ in this theory is the same as in the SM.

The NLO QCD-corrected Hamiltonian for $\Delta S=2$ transitions in the
minimal flavour violation supersymmetric framework has also been
obtained in Ref.~\cite{KS98}. From this, the result for $\epsilon$ can
be written as:
\begin{eqnarray}
\abseps &=& \frac{G_F^2f_K^2M_KM_W^2}{6\sqrt{2}\pi^2\Delta M_K}
\hat{B}_K\left(A^2\lambda^6\eta\right)
\bigl(y_c\left\{\hat{\eta}_{ct}f_3(y_c,y_t)-\hat{\eta}_{cc}\right\}
 \nonumber \\
&~& ~~~~~~~~~~~~~~~~~~~~+
~\hat{\eta}_{2}(K)SA^2\lambda^4(1-\rho)\bigr),
\label{eps2}
\end{eqnarray}
where the NLO QCD correction factor is \cite{KS98}:
\begin{eqnarray}
\hat{\eta}_{2}(K) &=& \alpha_s(m_c)^{\gamma^{(0)}/(2\beta_3^{(0)})}
\left(\frac{\alpha_s(m_b)}{\alpha_s(m_c)}\right)^{\gamma^{(0)}/(2\beta_4^{(0)})}
\left(\frac{\alpha_s(M_W)}{\alpha_s(m_b)}\right)^{\gamma^{(0)}/(2\beta_5^{(0)})} 
\nonumber\\
&& \left[ 1 + \frac{\alpha_s(m_c)}{4 \pi} (Z_3 - Z_4) +
\frac{\alpha_s(m_b)}{4 \pi} (Z_4 - Z_5) +
\frac{\alpha_s(M_W)}{4 \pi} (\frac{D}{S}+ Z_5) \right] ~.
\label{eta2k}
\end{eqnarray}
Here
\beq
Z_{n_f} = \frac{\gamma_{n_f}^{(1)}}{2 \beta_{n_f}^{(0)}} -
\frac{\gamma^{(0)}}{2 {\beta_{n_f}^{(0)}}^2} \beta_{n_f}^{(1)}~,
\eeq
and the quantities entering in Eqs.~(\ref{eta2s}) and (\ref{eta2k})
are the coefficients of the well-known beta function and anomalous
dimensions in QCD:
\begin{eqnarray}
\gamma^{(0)} &=& 6 \frac{N_c-1}{N_c}, ~~~ \beta_{n_f}^{(0)} = 
\frac{11N_c-2 n_f}{3} ~, \nonumber\\
\beta_{n_f}^{(1)} &=& \frac{34}{3}N_c^2 - \frac{10}{3} N_c n_f - 2 C_fn_f~,
\nonumber\\
\gamma_{n_f}^{(1)} &=& \frac{N_c-1}{2N_c} \left[-21 +\frac{57}{N_c} - 
\frac{19}{3}N_c + \frac{4}{3} n_f \right] ~,
\end{eqnarray}
with $N_c=3$ and $C_F=4/3$. The ratio
\beq 
\frac{\hat{\eta}_{2,S}(B)(NLO)}{\hat{\eta}_{2,S}(B)(LO)}
= 1 +\frac{\alpha_s(M_W)}{4\pi}
(\frac{D}{S} + Z_5),
\label{eta2lnl}
\eeq 
is worked out numerically in Ref.~\cite{KS98} as a function of the
supersymmetric parameters (chargino mass $m_{\chi_2}$, mass of the
lighter of the two stops $m_{\tilde{t}_R}$, and the mixing angle
$\phi$ in the stop sector). This ratio is remarkably stable against
variations in the mentioned parameters and is found numerically to be
about $0.89$. Since in the LO approximation the QCD correction factor
${\hat \eta}_{2,S}(B)(LO)$ is the same in the SM and SUSY, with
\beq
{\hat\eta}_{2,S}(B)(LO) = \alpha_s(M_W)^{\gamma^{(0)}/2\beta_{n_f}^{(0)}} ~,
\eeq
the QCD correction factor ${\hat \eta}_{2,S}(B)(NLO)$ entering in the
expressions for $\Delta M_d$ and $\Delta M_s$ in the MSSM is found to
be ${\hat \eta}_{2,S}(B)(NLO) = 0.51$ in the $\overline{MS}$-scheme. This is 
to be compared with the
corresponding quantity ${\hat \eta}_B = 0.55$ in the SM. Thus, NLO
corrections in $\Delta M_d$ (and $\Delta M_s$) are similar in the SM
and MSSM, but not identical.

The expression for ${\hat \eta}_{2,S}(K)(NLO)/{\hat
\eta}_{2,S}(K)(LO)$ can be expressed in terms of the ratio ${\hat
\eta}_{2,S}(B)(NLO)/{\hat \eta}_{2,S}(B)(LO)$ given above and the
flavour-dependent matching factors $Z_{n_f}$:
\beq
\frac{\hat{\eta}_{2,S}(K)(NLO)}{\hat{\eta}_{2,S}(K)(LO)}
= \frac{\hat{\eta}_{2,S}(B)(NLO)}{\hat{\eta}_{2,S}(B)(LO)} 
+ {\alpha_s(m_c) \over 4\pi} (Z_3 - Z_4)
+ {\alpha_s(m_b) \over 4\pi} (Z_4 - Z_5) \simeq 0.884 ~,
\eeq
where we have used the numerical value $\hat{\eta}_{2,S}(B)(NLO) /
\hat{\eta}_{2,S}(B)(LO) = 0.89$ calculated by Krauss and Soff
\cite{KS98}, along with $\alpha_s(m_c) = 0.34$ and $\alpha_s(m_b) =
0.22$. Using the expression for the quantity ${\hat
  \eta}_{2,S}(K)(LO)$, which is given by the prefactor multiplying the
square bracket in Eq.~(\ref{eta2k}), one gets ${\hat
  \eta}_{2,S}(K)(NLO) = 0.53$ in the $\overline{MS}$-scheme. This is
to be compared with the corresponding QCD correction factor in the SM,
${\hat\eta}_{tt} = 0.57$, given in Table \ref{datatable}. Thus the two
NLO factors are again very similar but not identical.

Following the above discussion, the SUSY contributions to $\delmd$,
$\delms$ and $\abseps$ in supersymmetric theories are incorporated in our
analysis in a simple form:
\begin{eqnarray}
\delmd &=& \delmd (SM) [ 1 +
f_d(m_{\chi_2^\pm},m_{\tilde{t}_R},
m_{H^\pm}, \tan \upsilon) ], \nonumber \\
\delms &=& \delms (SM) [ 1 +
f_s(m_{\chi_2^\pm},m_{\tilde{t}_R},
m_{H^\pm}, \tan \upsilon) ], \nonumber \\
\abseps &=& \frac{G_F^2f_K^2M_KM_W^2}{6\sqrt{2}\pi^2\Delta M_K}
\hat{B}_K\left(A^2\lambda^6\eta\right)
\bigl(y_c\left\{\hat{\eta}_{ct}f_3(y_c,y_t)-\hat{\eta}_{cc}\right\}
 \nonumber \\
&~& ~~~~~~~~~~~~~~+
~\hat{\eta}_{tt}y_tf_2(y_t)[1 + f_\epsilon
(m_{\chi_2^\pm},m_{\tilde{t}_2}, m_{H^\pm},
\tan \upsilon)] A^2\lambda^4(1-\rho)\bigr).
\label{susyformel}
\end{eqnarray}
The quantities $f_d$, $f_s$ and $f_\epsilon$ can be expressed as
\begin{eqnarray}
\label{fis}
f_d &=& f_s= \frac{\hat{\eta}_{2,S}(B)}{\hat{\eta}_B} R_{\Delta_d}(S),
\nonumber\\ f_\epsilon &=&
\frac{\hat{\eta}_{2,S}(K)}{\hat{\eta}_{tt}}R_{\Delta_d}(S),
\end{eqnarray}
where $R_{\Delta_d}(S)$ is defined as
\beq
R_{\Delta_d}(S) \equiv { \Delta M_d(SUSY) \over \Delta M_d(SM) } (LO)
= {S \over y_t f_2(y_t)} ~.
\eeq

The functions $f_{i}$, $i=d,s,\epsilon$ are all positive definite, i.e.\
the supersymmetric contributions add {\it constructively} to the SM 
contributions in the entire allowed supersymmetric parameter space. We
find that the two QCD correction factors appearing in Eq.~(\ref{fis})   
are numerically very close to one another, with
$\hat{\eta}_{2,S}(B)/\hat{\eta}_B \simeq
{\hat\eta}_{2,S}(K)/\hat{\eta}_{tt} = 0.93$. Thus, to an excellent
approximation, one has $f_d = f_s = f_\epsilon \equiv f$.

How big can $f$ be? This quantity is a function of the masses of the
top squark, chargino and the charged Higgs, $m_{\tilde{t}_R}$,
$m_{\tilde{\chi}^\pm_2}$ and $m_{H^\pm}$, respectively, as well as of
$\tan\upsilon$. The maximum allowed value of $f$ depends on the model
(minimal SUGRA, non-minimal SUGRA, MSSM with constraints from EDMs,
etc.).  We have numerically calculated the quantity $f$ by varying the
SUSY parameters $\phi$, $m_{\tilde{t}_R}$, $m_{\chi_2}$, $m_{H^\pm}$
and $\tan \upsilon$.  Using, for the sake of illustration,
$m_{\chi_2^\pm}=m_{\tilde{t}_R}=m_{H^\pm} =100$ GeV,
$m_{\chi_1^\pm}=400$ GeV and $\tan \upsilon=2$, and all other
supersymmetric masses much heavier, of $O(1)$ TeV, we find that the
quantity $f$ varies in the range:
\beq
 0.4 \leq f \leq 0.75 ~~~\mbox{for}~~~ \vert \phi \vert \leq \pi/4 ~,
\label{fnumber}
\eeq
with the maximum value of $f$ being at $\phi=0$.
These parametric values are allowed by the constraints from the NLO
analysis of the decay $B \to X_s + \gamma$ reported in
Ref.~\cite{CDGG98}, as well as from direct searches of the
supersymmetric particles \cite{PDG98}. The allowed range of
$f$ is reduced as $\tan \upsilon$ increases. Thus, for $\tan \upsilon=4$,
one has $0.15 \leq f \leq 0.42$ for $\vert \phi \vert \leq \pi/4$. 
Likewise, $f$ decreases as $m_{\tilde{t}_R}$,
$m_{\chi_2}$ and $m_{H^\pm}$ increase, though the dependence of $f$ on
$m_{H^\pm}$ is rather mild due to the compensating effect of the
$H^\pm$ and chargino contributions in the MSSM, as observed in
Ref.~\cite{CDGG98}. This sets the size of $f$ allowed by the present 
constraints in the minimal flavour violation version of the MSSM.

If additional constraints on the supersymmetry breaking parameters are
imposed, as is the case in the minimal and non-minimal versions of the
SUGRA models, then the allowed values of $f$ will be further
restricted. A complete NLO analysis of $f$ would require a
monte-carlo approach implementing all the experimental and theoretical
constraints (such as the SUGRA-type mass relations). In particular,
the NLO correlation between ${\cal B}(B \to X_s \gamma)$ and $f$ has
to be studied in an analogous fashion, as has been done, for example,
in Refs.~\cite{Gotoetal98-1,Gotoetal98-2} with the leading order SUSY
effects. 

In this paper we adopt an approximate method to constrain $f$ in
SUGRA-type models. We take the maximum allowed values of the quantity
$R_{\Delta_d}(S)$, defined earlier, from the existing LO analysis of
the same and obtain $f$ by using Eq.~(\ref{fis}). For the sake of
definiteness, we use the updated work of Goto et al.\ 
\cite{Gotoetal98-1,Gotoetal98-2}, which is based, among other
constraints, on the following:
\begin{itemize}

\item The lightest chargino mass is larger than $91$ GeV, and all other 
charged SUSY particle masses are larger than $80$ GeV.

\item The gluino and squark masses are bounded from the searches at TEVATRON
and LEP \cite{PDG98}. 

\item Constraints on the supersymmetric parameters taking into account
the NLO calculation of the decay branching ratio ${\cal B}(B\to X_s
\gamma)$ in the SM and the charged Higgs contribution in the MSSM
\cite{bsgSUSY}, and the updated experimental branching ratio from the
CLEO and ALEPH collaborations, which at 95\% C.L.\ is given by
\beq
2.0 \times 10^{-4} \le {\cal B}(B \to X_s \gamma) \le 4.5 \times 10^{-4} ~.
\eeq

\end{itemize}

This last constraint plays a rather crucial role in determining the
allowed values of $R_{\Delta_d}(S)$ --- and hence of $\Delta
M_d(SUSY)/\Delta M_d(SM)$ and $\abseps(SUSY)/\abseps(SM)$ --- as the 
magnitude of the 
SUSY contribution in $\Delta M_d$, $\abseps$ and ${\cal B}(B \to X_s 
\gamma)$ are strongly correlated for the small $\tan\upsilon$ case
\cite{Gotoetal98-1,Gotoetal98-2}. In fact, were it not for the bounds
on ${\cal B}(B \to X_s \gamma)$ given above, much larger values of
$R_{\Delta_d}(S)$ would be allowed. 

 From the published results we conclude that typically 
$f$ can be as large as $0.45$ in non-minimal SUGRA models for low $\tan
\upsilon$ (typically $\tan \upsilon=2$) \cite{Gotoetal98-1}, and
approximately half of this value in minimal SUGRA models
\cite{Nihei97,Gotoetal97,Gotoetal98-1}. Relaxing the SUGRA mass
constraints, admitting complex values of $A$ and $\mu$ but
incorporating the EDM constraints, and imposing the constraints mentioned
above,  Baek and Ko \cite{BK98-1} find that
$f$ could be as large as $f=0.75$. In all cases, the value of $f$
decreases with increasing $\tan \upsilon$ or increasing
$m_{\tilde{\chi}^\pm_2}$ and $m_{\tilde{t}_R}$, as noted above.

\subsection{SUSY Fits}

For the SUSY fits, we use the same program as for the SM fits, except
that the theoretical expressions for $\Delta M_d$, $\Delta M_s$ and
$\abseps$ are modified as in Eq.~(\ref{susyformel}). We compare the fits
for four representative values of the SUSY function $f$ --- 0, 0.2,
0.4 and 0.75 --- which are typical of the SM, minimal SUGRA models,
non-minimal SUGRA models, and non-SUGRA models with EDM constraints,
respectively.

\begin{figure}
\vskip -1.0truein
\centerline{\epsfxsize 3.5 truein \epsfbox {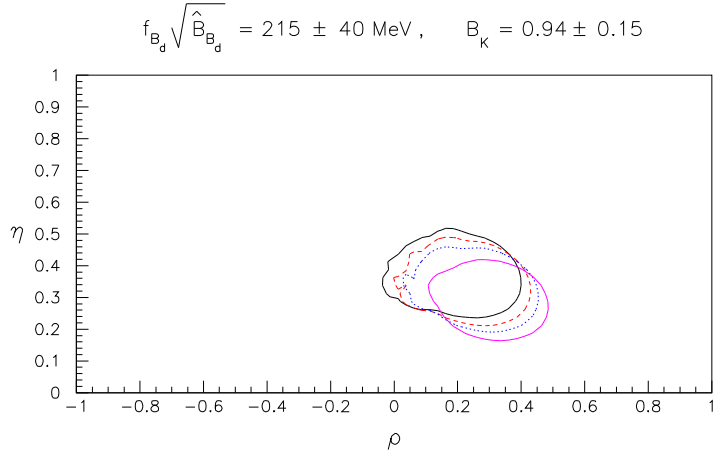}}
\vskip -1.4truein
\caption{Allowed 95\% C.L. region in $\rho$--$\eta$ space in the SM 
  and in SUSY models, from a fit to the data given in Table
  \protect{\ref{datatable}}. From left to right, the allowed regions
  correspond to $f=0$ (SM, solid line), $f=0.2$ (long dashed line),
  $f=0.4$ (short dashed line), $f=0.75$ (dotted line).}
\label{sugratot}
\end{figure}

The allowed 95\% C.L. regions for these four values of $f$ are all
plotted in Fig.~\ref{sugratot}. As is clear from this figure, there is
still a considerable overlap between the $f=0$ (SM) and $f=0.75$
regions. However, there are also regions allowed for one value of $f$
which are excluded for another value. Thus a sufficiently precise
determination of the unitarity triangle might be able to exclude
certain values of $f$ (including the SM, $f=0$).

{}From Fig.~\ref{sugratot} it is clear that a measurement of the CP
angle $\beta$ will {\it not} distinguish among the various values of
$f$: even with the naked eye it is evident that the allowed range
for $\beta$ is roughly the same for all models. Rather, it is the
measurement of $\gamma$ or $\alpha$ which has the potential to rule
out certain values of $f$. As $f$ increases, the allowed region moves
slightly down and towards the right in the $\rho$--$\eta$ plane,
corresponding to smaller values of $\gamma$ (or equivalently, larger
values of $\alpha$). We illustrate this in Table~\ref{cpasym1}, where
we present the allowed ranges of $\alpha$, $\beta$ and $\gamma$, as
well as their central values (corresponding to the preferred values of
$\rho$ and $\eta$), for each of the four values of $f$. From this
Table, we see that the allowed range of $\beta$ is largely insensitive
to the model. Conversely, the allowed values of $\alpha$ and $\gamma$
do depend somewhat strongly on the chosen value of $f$. Note, however,
that one is not guaranteed to be able to distinguish among the various
models: as mentioned above, there is still significant overlap among
all four models. Thus, depending on what values of $\alpha$ and
$\gamma$ are obtained, we may or may not be able to rule out certain
values of $f$.

One point which is worth emphasizing is the correlation of $\gamma$
with $f$. This study clearly shows that large values of $f$ require
smaller values of $\gamma$. The reason that this is important is as
follows. The allowed range of $\gamma$ for a particular value of $f$
is obtained from a fit to all CKM data, even those measurements which
are unaffected by the presence of supersymmetry. Now, the size of
$\gamma$ indirectly affects the branching ratio for $B \to X_s
\gamma$: a larger value of $\gamma$ corresponds to a smaller value of
$|V_{ts}|$ through CKM unitarity. And this branching ratio is among
the experimental data used to bound SUSY parameters and calculate the
allowed range of $f$. Therefore, the above $\gamma$--$f$ correlation
indirectly affects the allowed values of $f$ in a particular SUSY
model, and thus must be taken into account in studies which examine
the range of $f$. For example, it is often the case that larger values
of $f$ are allowed for large values of $\gamma$. However, as we have seen
above, the CKM fits disfavour such values of $\gamma$.

\begin{table}
\hfil
\vbox{\offinterlineskip
\halign{&\vrule#&
 \strut\quad#\hfil\quad\cr
\noalign{\hrule}
height2pt&\omit&&\omit&&\omit&&\omit&&\omit&\cr 
& $f$ && $\alpha$ && $\beta$ && $\gamma$ && $(\alpha,\beta,\gamma)_{\rm cent}$ & \cr
height2pt&\omit&&\omit&&\omit&&\omit&&\omit&\cr 
\noalign{\hrule}
height2pt&\omit&&\omit&&\omit&&\omit&&\omit&\cr 
& $f=0$ (SM) && $65^\circ$ -- $123^\circ$ && $16^\circ$ -- $35^\circ$ && 
$36^\circ$ -- $97^\circ$ && $(93^\circ, 25^\circ, 62^\circ)$ & \cr 
& $f=0.2$ && $70^\circ$ -- $129^\circ$ && $16^\circ$ -- $34^\circ$ && 
$32^\circ$ -- $90^\circ$ && $(102^\circ, 24^\circ, 54^\circ)$ & \cr 
& $f=0.4$ && $75^\circ$ -- $134^\circ$ && $15^\circ$ -- $34^\circ$ && 
$28^\circ$ -- $85^\circ$ && $(110^\circ, 23^\circ, 47^\circ)$ & \cr 
& $f=0.75$ && $86^\circ$ -- $141^\circ$ && $13^\circ$ -- $33^\circ$ && 
$23^\circ$ -- $73^\circ$ && $(119^\circ, 22^\circ, 39^\circ)$ & \cr 
height2pt&\omit&&\omit&&\omit&&\omit&&\omit&\cr 
\noalign{\hrule}}}
\caption{Allowed 95\% C.L. ranges for the CP phases $\alpha$, $\beta$ and 
  $\gamma$, as well as their central values, from the CKM fits in the
  SM $(f=0)$ and supersymmetric theories, characterized by the
  parameter $f$ defined in the text.}
\label{cpasym1}
\end{table}

For completeness, in Table~\ref{cpasym2} we present the corresponding
allowed ranges for the CP asymmetries $\sin 2\alpha$, $\sin 2\beta$
and $\sin^2 \gamma$. Again, we see that the allowed range of $\sin
2\beta$ is largely independent of the value of $f$. On the other hand,
as $f$ increases, the allowed values of $\sin 2\alpha$ become
increasingly negative, while those of $\sin^2 \gamma$ become smaller.

\begin{table}
\hfil
\vbox{\offinterlineskip
\halign{&\vrule#&
 \strut\quad#\hfil\quad\cr
\noalign{\hrule}
height2pt&\omit&&\omit&&\omit&&\omit&\cr
& $f$ && $\sin 2\alpha$ &&
$\sin 2\beta$ && $\sin^2 \gamma$ & \cr
height2pt&\omit&&\omit&&\omit&&\omit&\cr
\noalign{\hrule}
height2pt&\omit&&\omit&&\omit&&\omit&\cr
& $f=0$ (SM) && $-$0.91 -- 0.77 && 0.53 -- 0.94 && 0.35 -- 1.00 & \cr
& $f=0.2$ && $-$0.98 -- 0.65 && 0.52 -- 0.93 && 0.28 -- 1.00 & \cr
& $f=0.4$ && $-$1.00 -- 0.50 && 0.49 -- 0.93 && 0.22 -- 0.99 & \cr
& $f=0.75$ && $-$1.00 -- 0.14 && 0.45 -- 0.91 && 0.16 -- 0.91 & \cr
height2pt&\omit&&\omit&&\omit&&\omit&\cr
\noalign{\hrule}}}
\caption{Allowed 95\% C.L. ranges for the CP asymmetries $\sin 2\alpha$, $\sin
  2\beta$ and $\sin^2 \gamma$, from the CKM fits in the SM $(f=0)$ and
  supersymmetric theories, characterized by the parameter $f$ defined
  in the text.}
\label{cpasym2}
\end{table}

The allowed (correlated) values of the CP angles for various values of
$f$ can be clearly seen in Figs.~\ref{alphabetacorr} and
\ref{alphagammacorr}. As $f$ increases from 0 (SM) to 0.75, the change
in the allowed $\sin 2\alpha$--$\sin 2\beta$ (Fig.~\ref{alphabetacorr})
and $\alpha$--$\gamma$ (Fig.~\ref{alphagammacorr}) regions is quite
significant.

In Sec.~2.1, we noted that $|V_{ub}/V_{cb}|$, ${\hat B}_K$ and
$\fbd\sqrt{\hat{B}_{B_d}}$ are very important in defining the allowed
region in the $\rho$--$\eta$ plane. At present, these three quantities
have large errors, which are mostly theoretical in nature. Let us
suppose that our theoretical understanding of these quantities
improves, so that the errors are reduced by a factor of two, i.e.
\begin{eqnarray}
\left\vert { V_{ub} \over V_{cb} } \right\vert & = & 0.093 \pm 0.007 ~, \nn\cr
\hat{B}_K & = & 0.94 \pm 0.07 ~, \nn\cr
\fbd\sqrt{\hat{B}_{B_d}} & = & 215 \pm 20~{\rm MeV} ~.
\label{newdata}
\end{eqnarray}
How would such an improvement affect the SUSY fits?

We present the allowed 95\% C.L. regions ($f=0$, 0.2, 0.4, 0.75) for
this hypothetical situation in Fig.~\ref{sugratot_fut}. Not
surprisingly, the regions are quite a bit smaller than in
Fig.~\ref{sugratot}. More importantly for our purposes, the regions
for the different values of $f$ have become more separated from one
another. That is, although there is still a region where all four $f$
values are allowed, precise measurements of the CP angles have a
better chance of ruling out certain values of $f$.

\begin{figure}
\vskip -1.0truein
\centerline{\epsfxsize 3.5 truein \epsfbox {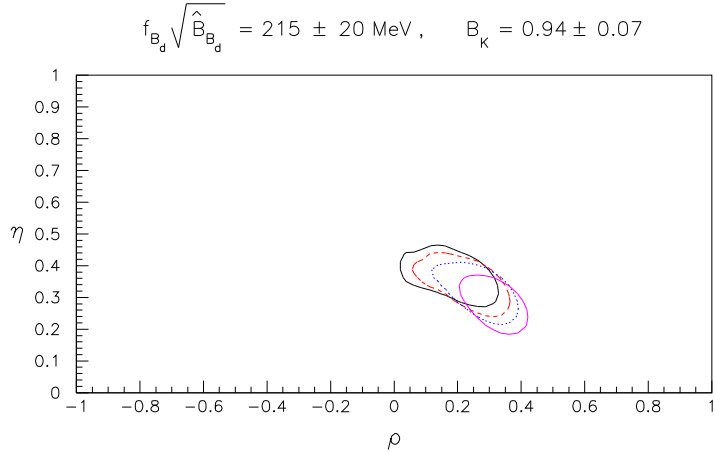}}
\vskip -1.4truein
\caption{Allowed 95\% C.L. region in $\rho$--$\eta$ space in the SM 
  and in SUSY models, from a fit to the data given in Table
  \protect{\ref{datatable}}, with the (hypothetical) modifications
  given in Eq.~(\protect\ref{newdata}). From left to right, the
  allowed regions correspond to $f=0$ (SM, solid line), $f=0.2$ (long
  dashed line), $f=0.4$ (short dashed line), $f=0.75$ (dotted line).}
\label{sugratot_fut}
\end{figure}

In Table~\ref{cpasym1_fut} we present the allowed ranges of $\alpha$,
$\beta$ and $\gamma$, as well as their central values, for this
scenario. Table~\ref{cpasym2_fut} contains the corresponding allowed
ranges for the CP asymmetries $\sin 2\alpha$, $\sin 2\beta$ and
$\sin^2 \gamma$. The allowed $\sin 2\alpha$--$\sin 2\beta$ and
$\alpha$--$\gamma$ correlations are shown in
Figs.~\ref{alphabetacorrfut} and \ref{alphagammacorrfut},
respectively. As is consistent with the smaller regions of
Fig.~\ref{sugratot}, the allowed (correlated) regions are
considerably reduced compared to Figs.~\ref{alphabetacorr} and
\ref{alphagammacorr}. As before, although the measurement of $\beta$
will not distinguish among the various values of $f$, the measurement
of $\alpha$ or $\gamma$ may.

\begin{figure}
\vskip -2.0truein
\centerline{\epsfxsize 8.0 truein \epsfbox {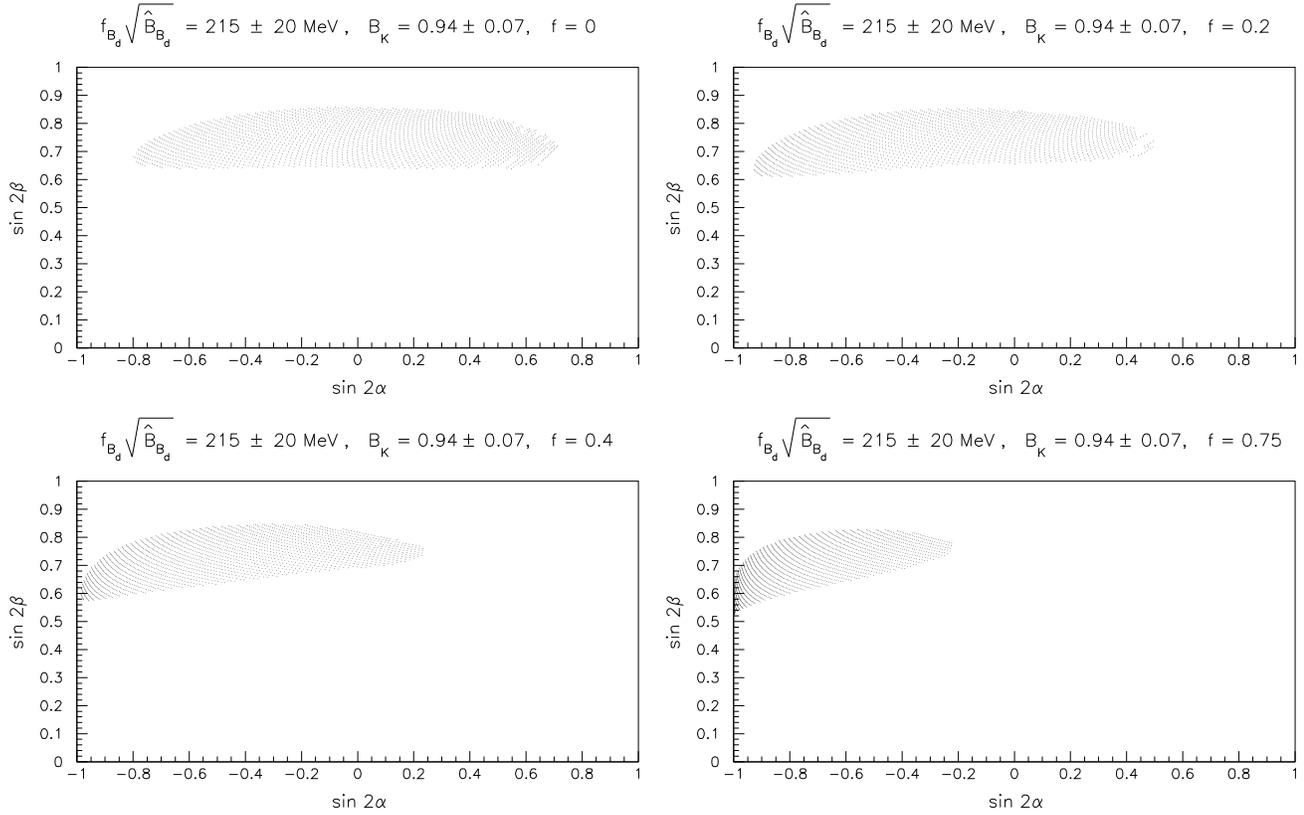}}
\vskip -4.7truein
\caption{Allowed 95\% C.L. region of the CP-violating quantities 
  $\sin 2\alpha$ and $\sin 2\beta$, from a fit to the data given in
  Table \protect{\ref{datatable}}, with the (hypothetical)
  modifications given in Eq.~(\protect\ref{newdata}). The upper left
  plot ($f=0$) corresponds to the SM, while the other plots ($f=0.2$,
  0.4, 0.75) correspond to various SUSY models.}
\label{alphabetacorrfut}
\end{figure}

\begin{figure}
\vskip -2.0truein
\centerline{\epsfxsize 8.0 truein \epsfbox {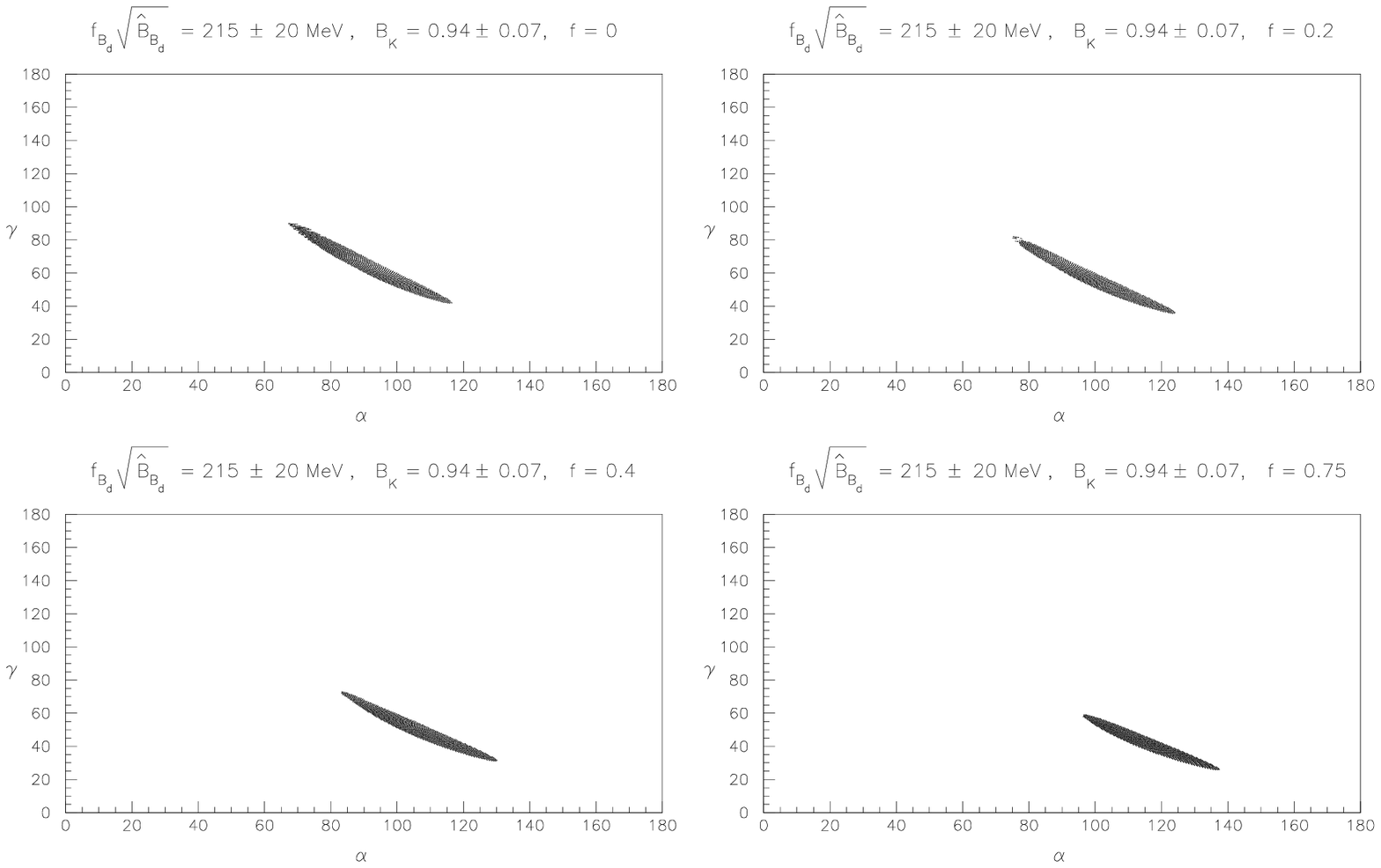}}
\vskip -4.7truein
\caption{Allowed 95\% C.L. region of the CP-violating quantities 
  $\alpha$ and $\gamma$, from a fit to the data given in Table
  \protect{\ref{datatable}}, with the (hypothetical) modifications
  given in Eq.~(\protect\ref{newdata}). The upper left plot ($f=0$)
  corresponds to the SM, while the other plots ($f=0.2$, 0.4, 0.75)
  correspond to various SUSY models.}
\label{alphagammacorrfut}
\end{figure}

Indeed, the assumed reduction of errors in Eq.~(\ref{newdata})
increases the likelihood of this happening. For example, consider
again Table~\ref{cpasym1}, which uses the original data set of
Table~\ref{datatable}. Here we see that $65^\circ \le \alpha \le
123^\circ$ for $f=0$ and $86^\circ \le \alpha \le 141^\circ$ for
$f=0.75$. Thus, if experiment finds $\alpha$ in the range
$86^\circ$--$123^\circ$, one cannot distinguish the SM ($f=0$) from
the SUSY model with $f=0.75$. However, consider now
Table~\ref{cpasym1_fut}, obtained using data with reduced errors.
Here, $67^\circ \le \alpha \le 116^\circ$ for $f=0$ and $97^\circ \le
\alpha \le 137^\circ$ for $f=0.75$. Now, it is only if experiment
finds $\alpha$ in the range $97^\circ$--$116^\circ$ that one cannot
distinguish $f=0$ from $f=0.75$. But this range is quite a bit smaller
than that obtained using the original data. This shows how an
improvement in the precision of the data can help not only in
establishing the presence of new physics, but also in distinguishing
among various models of new physics.

\begin{table}
\hfil
\vbox{\offinterlineskip
\halign{&\vrule#&
 \strut\quad#\hfil\quad\cr
\noalign{\hrule}
height2pt&\omit&&\omit&&\omit&&\omit&&\omit&\cr 
& $f$ && $\alpha$ && $\beta$ && $\gamma$ && $(\alpha,\beta,\gamma)_{\rm cent}$ & \cr
height2pt&\omit&&\omit&&\omit&&\omit&&\omit&\cr 
\noalign{\hrule}
height2pt&\omit&&\omit&&\omit&&\omit&&\omit&\cr 
& $f=0$ (SM) && $67^\circ$ -- $116^\circ$ && $20^\circ$ -- $30^\circ$ && 
$42^\circ$ -- $90^\circ$ && $(93^\circ, 24^\circ, 63^\circ)$ & \cr 
& $f=0.2$ && $74^\circ$ -- $124^\circ$ && $19^\circ$ -- $29^\circ$ && 
$36^\circ$ -- $82^\circ$ && $(102^\circ, 24^\circ, 54^\circ)$ & \cr 
& $f=0.4$ && $83^\circ$ -- $130^\circ$ && $18^\circ$ -- $29^\circ$ && 
$31^\circ$ -- $73^\circ$ && $(110^\circ, 23^\circ, 47^\circ)$ & \cr 
& $f=0.75$ && $97^\circ$ -- $137^\circ$ && $16^\circ$ -- $28^\circ$ && 
$26^\circ$ -- $59^\circ$ && $(119^\circ, 22^\circ, 39^\circ)$ & \cr 
height2pt&\omit&&\omit&&\omit&&\omit&&\omit&\cr 
\noalign{\hrule}}}
\caption{Allowed 95\% C.L. ranges for the CP phases $\alpha$, $\beta$ and 
  $\gamma$, as well as their central values, from the CKM fits in the
  SM $(f=0)$ and supersymmetric theories, characterized by the
  parameter $f$ defined in the text. We use the data given in Table
  \protect{\ref{datatable}}, with the (hypothetical) modifications
  given in Eq.~(\protect\ref{newdata}).}
\label{cpasym1_fut}
\end{table}

\begin{table}
\hfil
\vbox{\offinterlineskip
\halign{&\vrule#&
 \strut\quad#\hfil\quad\cr
\noalign{\hrule}
height2pt&\omit&&\omit&&\omit&&\omit&\cr
& $f$ && $\sin 2\alpha$ &&
$\sin 2\beta$ && $\sin^2 \gamma$ & \cr
height2pt&\omit&&\omit&&\omit&&\omit&\cr
\noalign{\hrule}
height2pt&\omit&&\omit&&\omit&&\omit&\cr
& $f=0$ (SM) && $-$0.80 -- 0.71 && 0.64 -- 0.86 && 0.44 -- 1.00 & \cr
& $f=0.2$ && $-$0.93 -- 0.53 && 0.61 -- 0.85 && 0.34 -- 0.98 & \cr
& $f=0.4$ && $-$0.99 -- 0.23 && 0.57 -- 0.85 && 0.27 -- 0.91 & \cr
& $f=0.75$ && $-$1.00 -- $-$0.23 && 0.52 -- 0.83 && 0.19 -- 0.73 & \cr
height2pt&\omit&&\omit&&\omit&&\omit&\cr
\noalign{\hrule}}}
\caption{Allowed 95\% C.L. ranges for the CP asymmetries $\sin 2\alpha$, $\sin
  2\beta$ and $\sin^2 \gamma$, from the CKM fits in the SM $(f=0)$ and
  supersymmetric theories, characterized by the parameter $f$ defined
  in the text. We use the data given in Table \protect{\ref{datatable}}, 
  with the (hypothetical) modifications given in 
Eq.~(\protect\ref{newdata}).} \label{cpasym2_fut}
\end{table}

\section{Conclusions}

In the very near future, CP-violating asymmetries in $B$ decays will
be measured at $B$-factories, HERA-B and hadron colliders. Such measurements
will give us crucial information about the interior angles $\alpha$,
$\beta$ and $\gamma$ of the unitarity triangle. If we are lucky, there
will be an inconsistency in the independent measurements of the sides
and angles of this triangle, thereby revealing the presence of new
physics.

If present, this new physics will affect $B$ decays principally
through new contributions to $B^0$--${\overline{B^0}}$ mixing. If these
contributions come with new phases (relative to the SM), then the CP
asymmetries can be enormously shifted from their SM values. In this
case there can be huge discrepancies between measurements of the
angles and the sides, so that the new physics will be easy to find.

In fact, there are several models which have new contributions, with
new phases, to $B^0$--${\overline{B^0}}$ mixing. However, these models
do not predict what those phases are. That is, there are no {\it
  predictions} for the values of CP asymmetries in such models. All
that can be said is that large effects are possible.

A more interesting possibility, from the point of view of making
predictions, are models which contribute to $B^0$--${\overline{B^0}}$
mixings and $\abseps$, but without new phases. One type of new physics which 
does
just this is supersymmetry (SUSY). There are some SUSY models which do
contain new phases, but they suffer from the problem described above:
lack of predictivity. However, there is also a large class of SUSY
models with no new phases. In this paper we have concentrated on these
models.

There has been an enormous amount of study of SUSY models over the
past two decades. Much of this work has concentrated on the minimal
supersymmetric standard model (MSSM), in which the new phases are
constrained by limits on the electric dipole moments of the neutron
and electron to be essentially zero. Taking into account supersymmetry
breaking, in supergravity (SUGRA) models, the SUSY-breaking parameters
of the MSSM are assumed to take a simple form at the Planck scale.
However, there are a variety of ways to do this, so that in fact there
is a fairly large class of SUSY models in which there are new
contributions to $B^0$--${\overline{B^0}}$ mixing, but no new phases.

In this paper we have examined the predictions of such models for the
unitarity triangle and explored the extent to which this type of new
physics can be discovered through measurements of the sides and angles
of the unitarity triangle.

In these models, there are new, supersymmetric contributions to
\kkbar, \bdbdbar\ and \bsbsbar\ mixing. The key ingredient in our
analysis is the fact that these contributions, which add
constructively to the SM, depend on the SUSY parameters in essentially
the same way. That is, so far as an analysis of the unitarity triangle
is concerned, there is a single parameter, $f$, which characterizes
the various SUSY models within this class of models ($f=0$ corresponds
to the SM). For example, the values $f=0.2$, 0.4 and 0.75 are found in
minimal SUGRA models, non-minimal SUGRA models, and non-SUGRA models
with EDM constraints, respectively.

We have therefore updated the profile of the unitarity triangle in
both the SM and some variants of the MSSM. We have used the latest 
experimental 
data on $|V_{cb}|$, $|V_{ub}/V_{cb}|$, $\Delta M_d$ and $\Delta M_s$, as
well as the latest theoretical estimates (including errors) of
$\hat{B}_K$, $\fbd\sqrt{\hat{B}_{B_d}}$ and $\xi_s \equiv
\fbd\sqrt{\hat{B}_{B_d}}/\fbs\sqrt{\hat{B}_{B_s}}$. In addition to
$f=0$ (SM), we considered the three SUSY values of $f$: 0.2, 0.4 and
0.75.

We first considered the profile of the unitarity triangle in the SM,
shown in Fig.~\ref{rhoeta1}. At present, the allowed ranges for the CP
angles at 95\% C.L. are
\beq
65^\circ \le \alpha \le 123^\circ ~~,~~~~
16^\circ \le \beta \le 35^\circ ~~,~~~~
36^\circ \le \gamma \le 97^\circ ~~,
\eeq
or equivalently,
\beq
-0.91 \le  \sin 2\alpha  \le 0.77 ~~,~~~~
0.52  \le  \sin 2\beta  \le 0.94  ~~,~~~~
0.35  \le  \sin^2 \gamma  \le 1.00 ~~.
\eeq

We then compared the SM with the different SUSY models. The result can
be seen in Fig.~\ref{sugratot}. As $f$ increases, the allowed region
moves slightly down and to the right in the $\rho$--$\eta$ plane. The
main conclusion from this analysis is that the measurement of the CP
angle $\beta$ will not distinguish among the SM and the various SUSY
models -- the allowed region of $\beta$ is virtually the same in all
these models. On the other hand, the allowed ranges of $\alpha$ and
$\gamma$ do depend on the choice of $f$. For example, larger values of
$f$ tend to favour smaller values of $\gamma$. Thus, with measurements
of $\gamma$ or $\alpha$, we may be able to rule out certain values of
$f$ (including the SM, $f=0$). However, we also note that there is no
guarantee of this happening -- at present there is still a significant
region of overlap among all four models.

Finally, we also considered a hypothetical future data set in which
the errors on $|V_{ub}/V_{cb}|$, $\hat{B}_K$ and
$\fbd\sqrt{\hat{B}_{B_d}}$, which are mainly theoretical, are reduced
by a factor of two. For two of these quantities ($|V_{ub}/V_{cb}|$ and
$\fbd\sqrt{\hat{B}_{B_d}}$), this has the effect of reducing the
uncertainty on the sides of the unitarity triangle by the same factor.
The comparison of the SM and SUSY models is shown in
Fig.~\ref{sugratot_fut}. As expected, the allowed regions for all
models are quite a bit smaller than before. Furthermore, the regions
for different values of $f$ have become more separated, so that
precise measurements of the CP angles have a better chance of ruling
out certain values of $f$.

\bigskip
\noindent
{\bf Acknowledgements}:
\bigskip

We would like to especially thank Seungwon Baek, John Ellis, Toru Goto
and Pyungwon Ko for very helpful discussions, and
Laksana Tri Handoko for numerical help.
Thanks are also due to Hans-Gunther Moser, Olivier Schneider and
Fabrizio Parodi for providing us the experimental information on the LEP
measurements of the $B$-$\overline{B}$ mixings and instructions in 
implementing the resulting constraints. D.L. is grateful to G. Azuelos 
and I. Trigger for their invaluable help with MINUIT and PAW. This 
research was financially supported by NSERC of Canada and FCAR du Qu\'ebec.

Note added: As we were submitting this paper for publication, we noted
the preprint by F. Parodi, P. Roudeau and A. Stocchi (LAL 99-03,
DELPHI 99-27 CONF 226, hep-ex/9903063), in which the constraints on the
parameters of the CKM matrix are presented in the SM. Though our input
parameters differ, in particular the assumed errors, our results in the 
SM are similar to theirs but not identical. 

\end{document}